\documentclass[compsoc]{IEEEtran}
\usepackage{amsmath,amsfonts}
\usepackage{algorithmic}
\usepackage{algorithm}
\usepackage{array}
\usepackage[caption=false,font=normalsize,labelfont=sf,textfont=sf]{subfig}
\usepackage{textcomp}
\usepackage{stfloats}
\usepackage{url}
\usepackage{verbatim}
\usepackage{graphicx}
\usepackage{cite}

\usepackage{booktabs, multirow} 
\usepackage{soul}
\usepackage[table]{xcolor} 
\usepackage{changepage,threeparttable} 
\usepackage{amssymb}
\usepackage{longtable} 
\usepackage{pdflscape}
\usepackage{afterpage}
\usepackage{xspace}
\usepackage{wrapfig}

\newcommand{\dr}{dimensionality reduction\xspace}
\newcommand{\drshort}{DR\xspace}
\newcommand{\drdata}{dimensionally-reduced data\xspace}

\makeatletter
 \def\SOUL@hlpreamble{%
 \setul{}{2ex}
 \let\SOUL@stcolor\SOUL@hlcolor
 \SOUL@stpreamble
 }
\makeatother

\definecolor{darkgray}{rgb}{0.31, 0.31, 0.31}

\definecolor{lightgray}{rgb}{0.87, 0.87, 0.87}

\definecolor{myblue}{RGB}{141,160,203}
\definecolor{mypink}{RGB}{231,138,195}
\definecolor{mylgreen}{RGB}{166,216,84}

\newenvironment{catfontenv}{\fontfamily{pag}\selectfont}{}
\DeclareRobustCommand{\catfont}[1]{{\begin{catfontenv}#1\end{catfontenv}}}

\usepackage{enumitem}

\usepackage{longfbox} 

\makeatletter
\newdimen\@tempdimd
\makeatother

\newfboxstyle{patternparam}{padding-top=2pt, padding-bottom=3pt, padding-left=3pt, padding-right=2pt, border-style=none, height=6.5pt, border-radius=4pt}

\newfboxstyle{patternparambox}{padding-top=2pt, padding-bottom=3.5pt, padding-left=3pt, padding-right=2pt, border-style=none, height=6.5pt}

\newcommand{\bluebox}[1]{\lfbox[patternparam, background-color=myblue]{{\color{white}{{\catfont{#1}}}}}}

\newcommand{\graybox}[1]{\lfbox[patternparambox, background-color=lightgray]{{{{\catfont{#1}}}}}}

\def\rqone{\bluebox{RQ1}}
\def\rqtwo{\bluebox{RQ2}}
\def\rqthree{\bluebox{RQ3}}
\def\rqfour{\bluebox{RQ4}}

\newcommand{\paran}[1]{(\hspace{1pt}#1\hspace{1pt})\xspace}

\DeclareRobustCommand{\bio}{\graybox{Biology}\xspace}
\DeclareRobustCommand{\chem}{\graybox{Chemistry}\xspace}
\DeclareRobustCommand{\phys}{\graybox{Physics}\xspace}
\DeclareRobustCommand{\busi}{\graybox{Business}\xspace}

\usepackage{hyperref}

\begin{document}

\title{A Critical Analysis of the Usage of Dimensionality Reduction in Four Domains}
\author{Dylan Cashman, Mark Keller, Hyeon Jeon, Bum Chul Kwon, Qianwen Wang
\IEEEcompsocitemizethanks{
\IEEEcompsocthanksitem
Dylan Cashman is with Brandeis University.\\ E-mail: dylancashman@brandeis.edu

\IEEEcompsocthanksitem
Mark Keller is with Harvard Medical School.\\ E-mail: mark\_keller@hms.harvard.edu

\IEEEcompsocthanksitem
Hyeon Jeon is with Seoul National University.\\ E-mail: hj@hcil.snu.ac.kr

\IEEEcompsocthanksitem
Bum Chul Kwon is with IBM Research.\\ E-mail: bumchul.kwon@us.ibm.com

\IEEEcompsocthanksitem
Qianwen Wang is with University of Minnesota.\\ E-mail: qianwen@umn.edu
}
}

\markboth{PREPRINT PUBLISHED JULY 14 2025.  To be published in IEEE Transactions on Visualization and Computer Graphics in 2025}%
{Cashman \MakeLowercase{\textit{et al.}}: Critical Analysis of Dim Red}

\IEEEpubid{DOI 10.1109/TVCG.2025.3567989 ~\copyright~2025 IEEE}

\IEEEtitleabstractindextext{
\begin{abstract}
Dimensionality reduction is used as an important tool for unraveling the complexities of high-dimensional datasets in many fields of science, such as cell biology, chemical informatics, and physics. Visualizations of the \drdata enable scientists to delve into the intrinsic structures of their datasets and align them with established hypotheses. Visualization researchers have thus proposed many dimensionality reduction methods and interactive systems designed to uncover latent structures.  At the same time, different scientific domains have formulated guidelines or common workflows for using dimensionality reduction techniques and visualizations for their respective fields. In this work, we present a critical analysis of the usage of dimensionality reduction in scientific domains outside of computer science.  First, we conduct a bibliometric analysis of 21,249 academic publications that use dimensionality reduction to observe differences in the frequency of techniques across fields.  Next, we conduct a survey of a 71-paper sample from four fields: biology, chemistry, physics, and business.  Through this survey, we uncover common workflows, processes, and usage patterns, including the mixed use of confirmatory data analysis to validate a dataset and projection method and exploratory data analysis to then generate more hypotheses.  We also find that misinterpretations and inappropriate usage is common, particularly in the visual interpretation of the resulting dimensionally reduced view.  Lastly, we compare our observations with recent works in the visualization community in order to match work within our community to potential areas of impact outside our community.  By comparing the usage found within scientific fields to the recent research output of the visualization community, we offer both validation of the progress of visualization research into dimensionality reduction and a call for action to produce techniques that meet the needs of scientific users.
\end{abstract}

}

\maketitle

\section{Introduction}
\label{sec:intro}

Recent years have witnessed the ubiquitous application of \dr (\drshort) techniques across various domains.
By converting hundreds or thousands of dimensions into just two, \drshort enables intuitive visualization of high-dimensional data, which greatly aids exploratory data analysis (EDA), especially in noise filtering and pattern identification. 
Consequently, \drshort has evolved into an essential component of many data analysis workflows \cite{sacha2016visual, nonato18multidimensional}.
One of the most popular \drshort methods, t-SNE~\cite{van2008tsne}, has been cited more than 45,000 times as of June 2024, according to Google Scholar.  In addition, \drshort is a core component in many visual analytics systems~\cite{cheng23polyphony, narechania22vitality, li24visual}.

At the same time, concerns regarding the use and interpretation of these techniques have emerged~\cite{chari2023specious, aupetit2007visualizing, lause24plos, jeon2025unveiling}.
First, distortion and information loss are inevitably induced when applying \drshort can easily lead to inaccurate interpretation of the high-dimensional data \cite{nonato18multidimensional, aupetit2007visualizing, jeon22measuring}. 
Second, inexperienced users often interpret projections in a manner that violates their original design intention.
For example, one common mistake is the assumption that the distance between two clusters in the projected 2D manifold directly reflects the similarity between them~\cite{wattenberg2016use, jeon24classes}.
There is ongoing debate regarding the usage and interpretation of dimensionality reduction in data analysis.

It is important for visualization researchers to consider if the visualization of \drshort results is meeting the needs of the scientific community at large.  
While there are tools and preprocessing pipelines recommended within domains, they may be designed explicitly for particular types and scales of data typical in those domains and thus not be more broadly applicable.  
Despite the ubiquitous use of \drshort, there does not appear to be a standard workflow for interpretation and use across domains.  Different users and fields approach the analysis in unique ways.    There may be a gap in the needs of the scientific community and the current capabilities of \drshort algorithms and visualization systems that integrate them.


To address these challenges,
we present a critical analysis of the usage of \drshort for high dimensional data analysis \textit{outside of computer science}.
We seek to answer the following research questions.

\begin{itemize}[leftmargin=10.1mm]
    \setlength\itemsep{4.5pt}
    \item [\rqone] \textbf{What are the usage patterns of \drshort outside of computer science?}  These usage patterns may differ by what particular techniques they are using (t-distributed Stochastic Neighbor Embedding i.e. t-SNE, Principal Component Analysis i.e. PCA, or Uniform Manifold Approximation and Projection i.e. UMAP) or by what characterizes the data that is used.  Answering this question informs the algorithmic design of visualization systems that allow for the exploration of high-dimensional data.
    \item [\rqtwo] \textbf{How are views of \drdata interpreted?}  The methods of interpretation may also depend on the analytic goals and what tasks are used to reach those analytic goals.  
    \item [\rqthree] \textbf{Do the usage patterns and interpretation methods differ by field?}  The variance in usage and interpretation across domains can provide insight into the generalizability and extensibility required by \drshort algorithms and tools.
    \item [\rqfour] \textbf{Are there any gaps in the available \drshort algorithms or tools that are opportunities for visualization researchers to adapt existing tools or to develop new tools?}  There may be low-hanging fruit to apply the learnings from design studies in visualization research to meet unmet needs within domain use cases that haven't been addressed adequately by our community.
\end{itemize}

To answer these research questions, we follow a three-step analysis of scientific literature.  First, we conduct a bibliometric analysis of 21,249 academic publications within scientific domains that use \drshort (\autoref{sec:bibliometrics}). The analysis determined broad trends and practices \paran{\rqone} and how those practices differ between communities at a high level \paran{\rqthree}.  Next, we present a survey of 71 papers from four scientific fields (\bio, \chem, \phys, and \busi) to uncover lower-level differences between usage patterns \paran{\rqone} and the interpretations of dimensionally-reduced views of data offered within scientific publications \paran{\rqtwo} (\autoref{sec:full-analysis}).  Full results of our survey are presented in a table in the appendix as well as an online browser of screenshots and metadata available at \url{https://dimension-reduction-vis.github.io/}.

We summarize the results from these two steps of the investigation in a comparative analysis of the findings from all fields in \autoref{sec:findings}.  
As part of our analysis of these results, we review concurrent literature within the visualization community, including STARs, surveys, and reviews.  These works typically review the usage of \drshort within visualization systems, and we conclude that there is a mismatch between the tasks identified in visualization systems vs. those found in our sample of domain science papers.
Through this comparison, we uncover gaps between the visualization needs in domain scientists and the solutions offered by visualization researchers.  We describe how these gaps lead to opportunities for visualization researchers \paran{\rqfour}.

We find that simple linear projections like PCA are much more frequently used than nonlinear techniques like t-SNE and UMAP, even though nonlinear techniques are more frequently found in visual analytics systems~\cite{espadoto2021toward}.  We also find that \dr visualizations are frequently used for \textit{both} confirmatory and exploratory data analysis, in ways that may lead to bias or spurious interpretations.  We describe three common workflows across the spectrum of confirmatory to exploratory data analysis.  We outline the different ways that the visualization of the dimensionally reduced data is interpreted across four fields.  In section~\ref{sec:discussion}, we provide takeaways to domain scientists and visualization researchers, including open questions for how visualization researchers can provide simpler-to-use projection techniques that provide insights on the underlying data while mitigating bias.


\section{Related Work}

\label{sec:relatedwork}

In this section, we review related surveys, state-of-the-art reports, and other publications that investigate the usage of \drshort within visualization, our four selected subject areas (biology, chemistry, physics, and business), and science at large.  
Our goal in this section is to situate our work against existing studies so that a reader might know first what novel findings are presented in this work, and second where to look for alternative viewpoints on the topic.  

We emphasize that our survey is the \textit{first} survey from a visualization researcher's viewpoint looking outwardly at the usage of \drshort in domain sciences. Our survey thus differentiates from existing surveys that look inwardly at the usage of \drshort within visualization research \cite{sacha2016visual, nonato18multidimensional} and surveys of domain scientists looking inwardly at the usage of \drshort within their own domains \cite{becht2019dimensionality, ujas2023guide, clarke2021tutorial, anders2018dissecting}.  Here, we review both to contrast their methods from this critical analysis.

\subsection{Surveys on \drshort from Visualization Researchers}

Visualizing and interacting with \drshort methods have become important topics in the visualization community, sparking visualization researchers to conduct various surveys.  We categorize previous literature into the target of their analysis.

\noindent \textbf{Survey on DR Techniques}
The most common type of DR-related surveys is, not surprisingly, the ones about DR techniques. These surveys aim to clarify the advantages and disadvantages of DR techniques, thereby supporting practitioners in selecting proper DR techniques in their analysis \cite{engel2012survey, nonato18multidimensional, espadoto2021toward, etemadpour15tvcg}. 
For example, Espadoto et al. \cite{espadoto2021toward} surveyed 44 DR techniques and quantitatively examined their performance using five quality metrics. Nonato and Aupetit \cite{nonato18multidimensional} also compared 28 different techniques, providing guidelines to select DR techniques by the analytic task. Etemadpour et al. \cite{etemadpour15tvcg}, Xia et al. \cite{xia22tvcg}, and Sedlmair et al. \cite{sedlmair13tvcg} also provide similar guidelines, where empirical user studies ground their guides. 

\noindent \textbf{Survey on Tasks}
This family of surveys taxonomizes DR-related analytic tasks. They thereby aim to gain a further understanding of how practitioners use and interact with DR projections. 
For example, Sacha et al. surveyed visualization papers that included interaction techniques with dimensionality reduction algorithms, revealing the procedure in which analysts interact with DR~\cite{sacha2016visual}.  
Nonato and Aupetit \cite{nonato18multidimensional} systematically survey and taxonomize the type of analytic tasks. It is not a literature survey, but Brehmer et al. \cite{brehmer2014visualizing} also revealed task sequence using DR projections for HD data analysis by conducting interview studies with analysts. 

\noindent \textbf{Survey on DR Quality metrics}
Quality metrics for DR assess the extent to which DR projections suffer from distortions \cite{jeon23zadu}. As different quality metrics focus on different structural characteristics (e.g., local neighborhood structure \cite{Venna2010InformationRP} or cluster structure \cite{jeon22measuring}), selecting appropriate DR quality metrics that match target analytic tasks is important for reliable data analysis. Surveys regarding quality metrics, therefore, aim to guide analysts in choosing appropriate metrics. 
Thurn et al. \cite{thrun2023analyzing} and Bertini et al. \cite{bertini2011quality}, for example, organize DR quality metrics regarding which structural characteristics they focus on. Lee and Verleysen \cite{lee09neurocomp} share a similar goal but concentrate on neighborhood preservation-based methods.

Our work differs from these works by focusing on the usage patterns and interpretations of DR methods across diverse fields of science, rather than limiting our scope to the visualization community. While previous surveys potentially oversample from design studies driven by visualization researchers or collaborations including them, we aim to investigate whether the usage patterns \paran{\rqone} and interpretations \paran{\rqtwo} of DR in the wild differ from those familiar to visualization experts. Unlike prior analyses that compare and contrast various visualization works within the same field, we compare and contrast the application of DR methods across diverse scientific disciplines \paran{\rqthree}. By conducting extrinsic observations, our study provides insights that have the potential to guide the development of more generally applicable tools \paran{\rqfour}, especially for managing biases and distortions inherent in DR techniques.

\subsection{Surveys from Subject Area Researchers}

Within individual subject areas, surveys, meta-analyses, or guidelines, papers act as an implicit or explicit reference on how to use various types of algorithms as part of the analysis of that subject area's data.  Within biology, there exist several works comparing the usage of different dimensionality reduction algorithms on various types of biological data~\cite{becht2019dimensionality, ujas2023guide, clarke2021tutorial, kimball2018beginner, liechti2021updated}.  Similar works can be found in physics and astronomy~\cite{traven2017galah}, epidemiology~\cite{sakaue2020dimensionality}, and chemistry~\cite{anders2018dissecting}.  These works provide some review of practices within the silo of a single field, and present some recommendations on the usage of \drshort within a particular high-dimensional data analysis pipeline.  In contrast, our work looks both within and across domains.  We also present our findings from a computer science perspective, interrogating the gap between the usage of \dr in visualization research and those being used out in the wild.

\section{Bibliometric Analysis Across Subject Areas}
\label{sec:bibliometrics}


In this section, we use a bibliometric analysis of all recent papers citing the publications that introduce \drshort methods.  

\subsection{Bibliometric Analysis Objectives and Design}

This analysis aims to identify broad trends and practices across domains outside of computer science \paran{\rqone} and how those practices differ between communities at a high level \paran{\rqthree}.  

First, we identify 78 publications initially introducing DR methods ("\drshort papers"), based on Table 1 of a recent survey of \drshort methods by Espadoto et al. \cite{espadoto2021toward}.  We exclude from our analysis general machine learning methods such as ``neural networks'' which were considered in the original table from Espadoto et al. \cite{espadoto2021toward} but are used more generally for other methods than \drshort.  We also exclude methods for which a single originating publication to reference could not be identified.  We acknowledge that this may bias our analysis, but we believe that the 78 papers provide a broad enough sample to provide insights.
Using the Semantic Scholar Academic Graph \cite{kinney2023theSS}, we query for all academic papers (excluding preprints) published since 2013 that cite those publications.
We use the Semantic Scholar classification of subject area, which is calculated using a machine learning method (see Appendix for more details).  We exclude citing papers in Computer Science, Mathematics, and Engineering from our analysis based on the subject areas on record in Semantic Scholar, as we are interested in understanding how DR methods are used in other subject areas.

We present results in the form of counts and proportions.  However, proportions are influenced by the introduction of new \drshort methods: the proportion of t-SNE usage might drop as a result of UMAP being introduced and used \textit{in addition} to t-SNE).  This could result in an observed decrease in the proportional usage of t-SNE, even if its use consistently increased over the period of analysis.  As a result, we additionally compute percentile rankings of papers using the CP-EX method described by Bornmann and Williams \cite{bornmann2020evaluation}, using the entire Semantic Scholar corpus to build cumulative percentages of papers with each citation count value in each subject area and year.

Acknowledging that citation counts are influenced by field size, we also consider metrics that are standardized to enable comparison across subject areas.
In addition, we stratify citation counts by cited method and subject area and convert them to rankings.


\subsection{Bibliometric Analysis Results and Discussion}

\begin{figure}
    \centering
    \includegraphics[width=\linewidth]{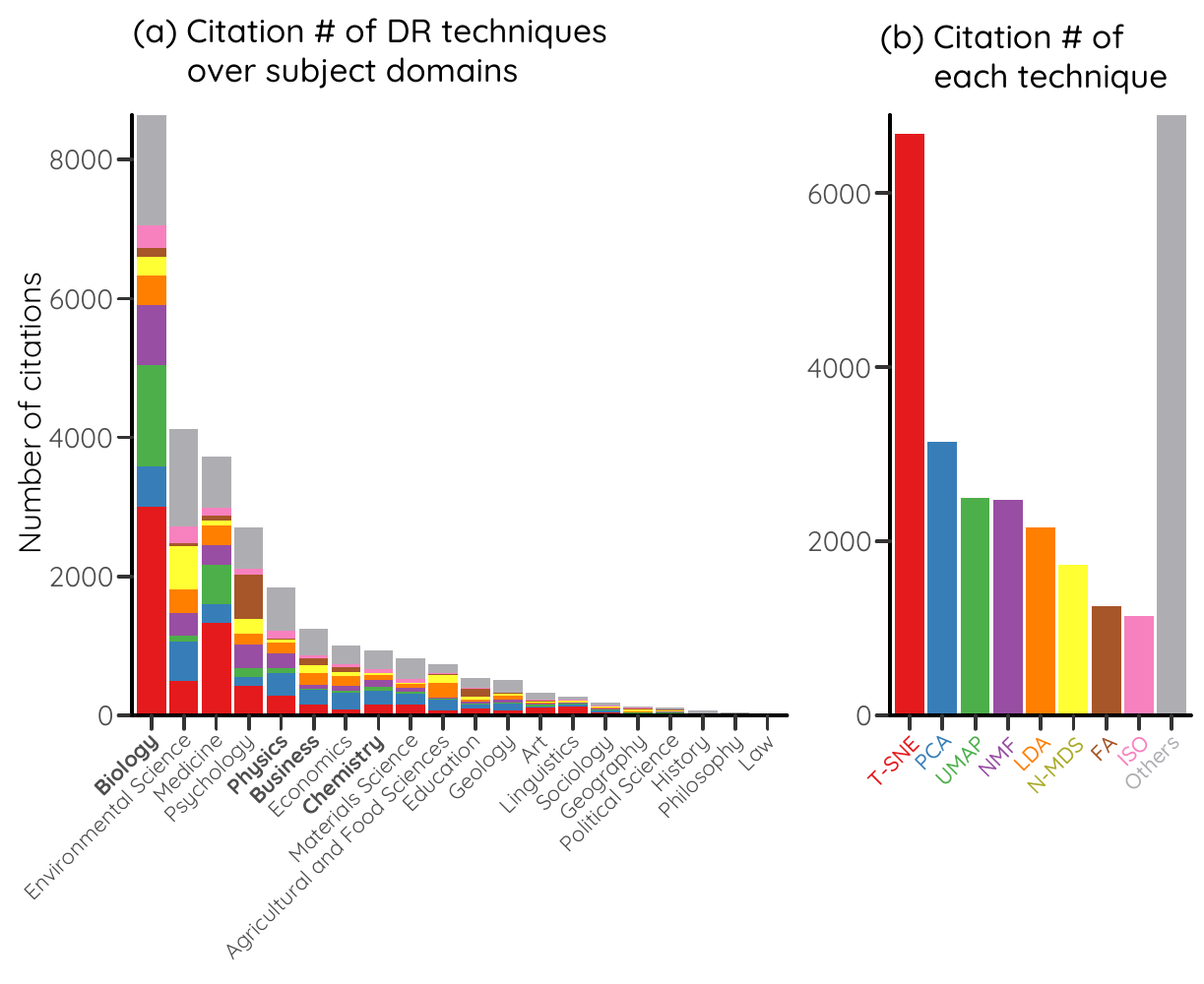}
    \vspace{-6mm}
    \caption{Results of our bibliometric analysis on citation counts.
    (a) The number of citations of DR techniques over each subject domain. (b) The number of citations in which each DR technique was obtained. }
    \label{fig:bib_count}
\end{figure}

\begin{figure}
    \centering
    \includegraphics[width=\linewidth]{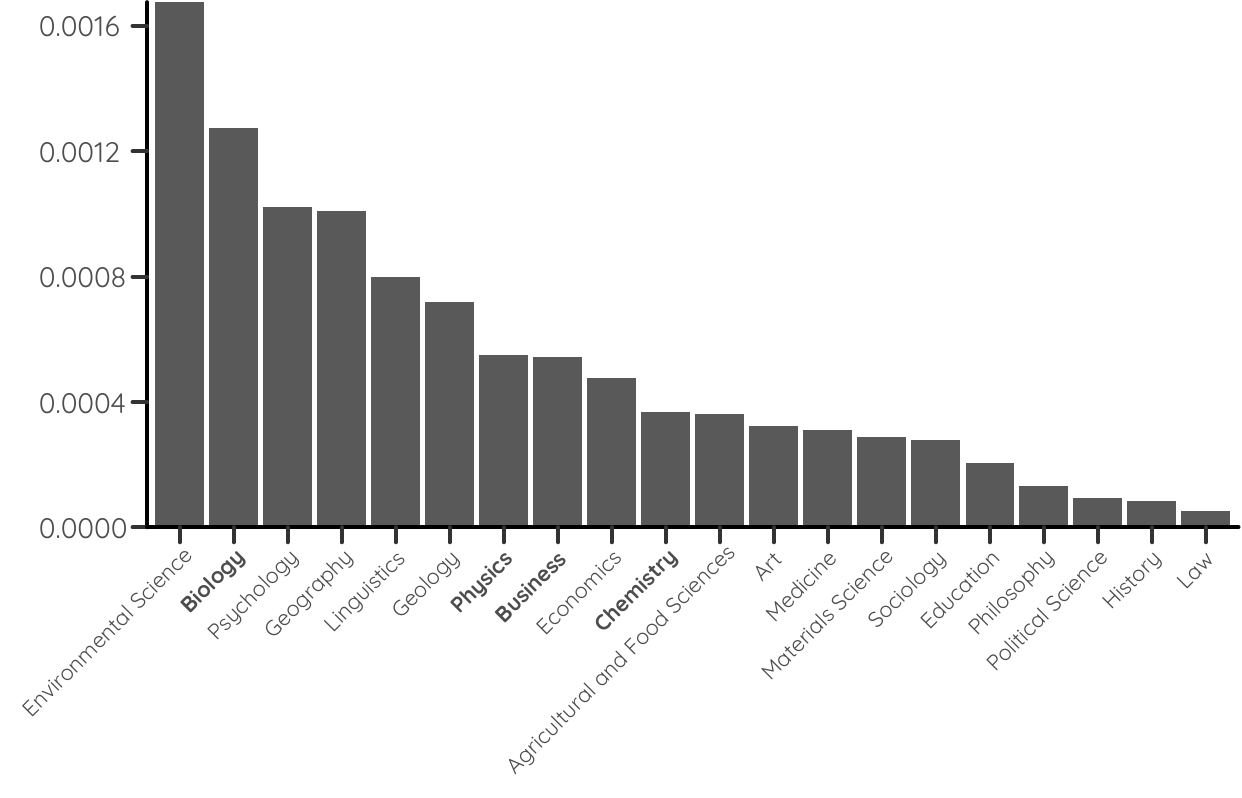}
    \vspace{-6mm}
    \caption{Results of our bibliometric analysis. The Y axis encodes the fraction of the number of papers in each subject area that cite a DR paper over the total number of published papers in the subject area.}
    \label{fig:bib_fraction}
\end{figure}

We first sought to quantify the usage of the 78 DR methods within each subject area based on direct citations.
We find that there are 136,956 unique papers published between 2013 and 2023 that cite at least one of the DR papers in the dataset.
Of those 136,956 papers, 74,720 have been mapped to at least one subject area via one of the two subject area mapping methods.
74,357 papers (99.5\% of those with at least one subject area) have been mapped to at most three subject areas using either method.
Using the latter subject area assignments, there are 21,249 citing papers assigned to at least one of 20 subject areas other than Computer Science, Mathematics, and Engineering.
We use these 21,249 papers and subject areas for the results that follow in this section. We will highlight the four scientific domains that we focus on in our subsequent literature review to call attention to the variance in the domains that we have chosen.

Through summation of citing paper counts within subject areas, we observe that the top ten areas citing the 78 DR papers are \bio (first), Environmental Science, Medicine, Psychology, \phys, \busi, Economics, \chem, Materials Science, and Agricultural and Food Sciences (tenth) (\autoref{fig:bib_count}a).
Instead of summing across subject areas, we find that t-SNE, PCA, and UMAP are the three most highly cited methods (\autoref{fig:bib_count}b).

\begin{figure*}[ht]
\centering
\includegraphics[width=\textwidth]{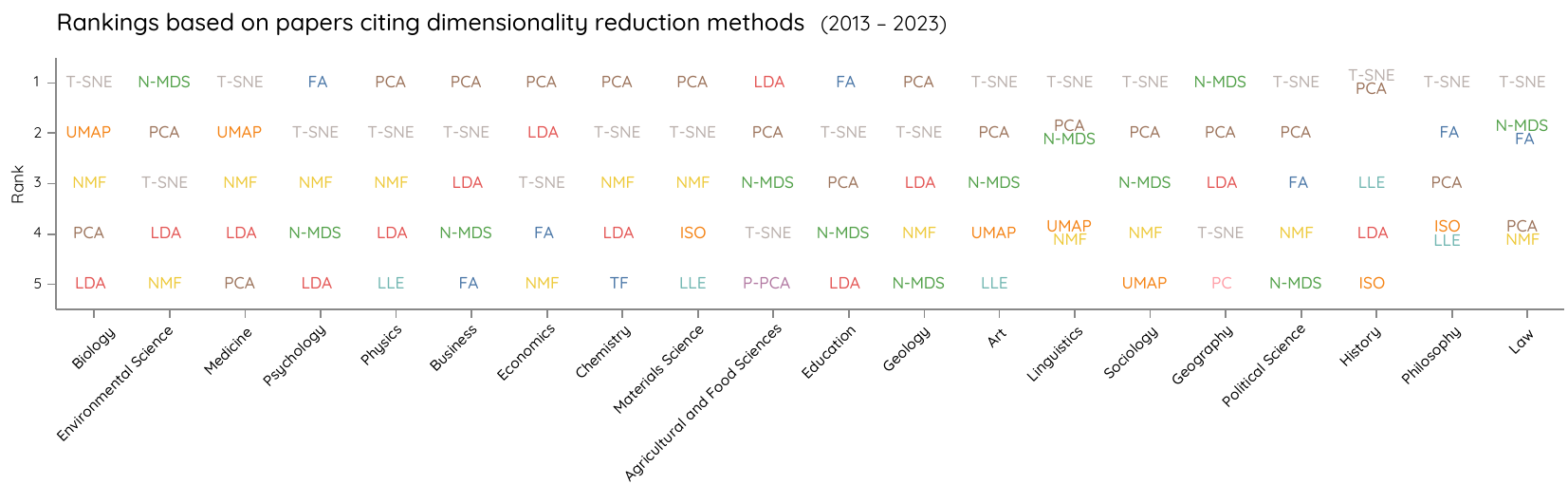}
\vspace{-7mm}
\caption{Rankings based on papers citing dimensionality reduction methods (2013-2023). Methods are ranked based on the number of citing papers in each subject area. Subject areas are ordered left-to-right based on descending total count of papers citing the 78 considered dimensionality reduction methods. Ties are displayed in alphabetical order. Full method names can be found in the appendix.}
\label{fig:bibplot_ranking_text}
\end{figure*}

Considering the number of citing papers as a proportion of all papers in each subject area, we instead see that the top ten areas are Environmental Science (one in 596 papers cites a DR method), \bio, Psychology, Geography, Linguistics, Geology, \phys, \busi, Economics, and \chem (one in 2,722 papers cites a DR method) (\autoref{fig:bib_fraction}).
When we stratify citation counts by cited method and subject area and convert them to rankings, it becomes apparent that while t-SNE and PCA appear in the top three methods for 18 and 16 of the 20 subject areas, respectively, UMAP is only within the top three methods for Biology and Medicine (\autoref{fig:bibplot_ranking_text}).
Other notable findings are that N-MDS is the top-ranked method in Environmental Science and Geography, FA is the top-ranked method in Psychology and Education, and LDA is the top-ranked method in Agricultural and Food Sciences (\autoref{fig:bibplot_ranking_text}). 

\begin{figure*}[ht]
\centering
\includegraphics[width=\textwidth]{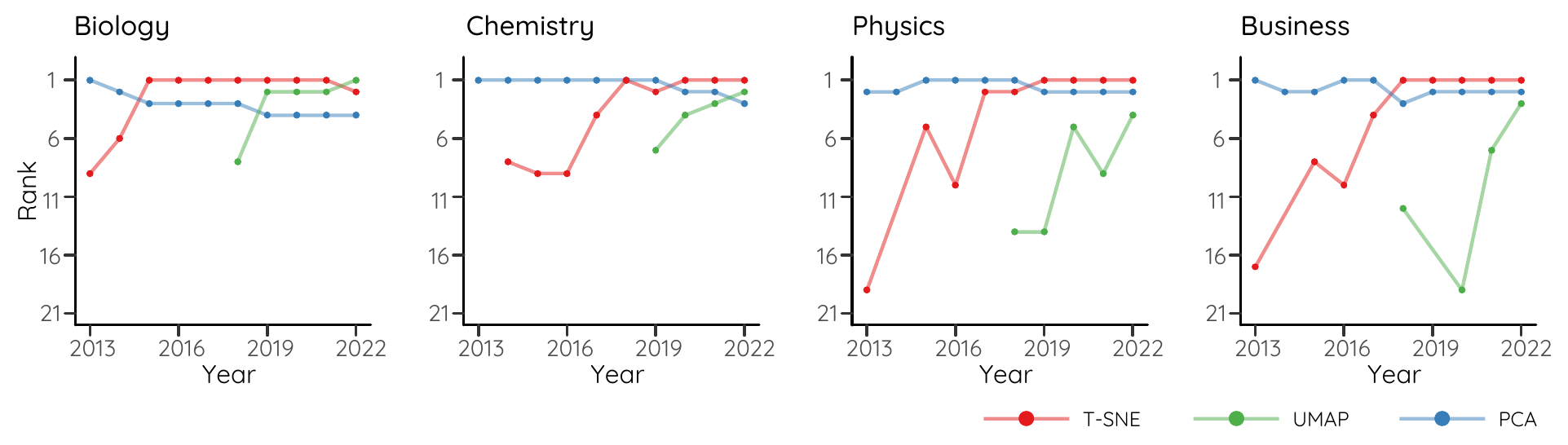}
\vspace{-7mm}
\caption{Percentage rankings of three most widely used DR techniques (T-SNE, UMAP, and PCA) across years (2013 -- 2022) in four domains we focused on. While PCA's percentage ranking stayed still or slightly decreased, the ranking of UMAP and t-SNE drastically increased through the past decade.}
\label{fig:bibplot_time}
\end{figure*}

For ease of interpretation and visualization, we next consolidate lesser-cited DR methods by mapping those that do not appear in the top three methods cited within any subject area to the category `Other'.
This results in nine DR method categories: t-SNE, PCA, UMAP, Nonnegative Matrix Factorization (NMF), Linear Discriminant Analysis (LDA), Non-metric Multidimensional Scaling (N-MDS), Factor Analysis (FA), Locally Linear Embedding (LLE), and `Other'.
Using these nine top-used DR method categories, we next considered the percent usage of each DR method within each subject area.
We find that the percentage of citations mapped to the `Other' method category is largest (compared to the eight remaining categories) in the subject areas Environmental Science, \phys, \busi, Economics, \chem, Materials Science, Education, Geology, History, and Philosophy, which suggests that there is not a single overwhelmingly dominant DR method used in these fields.
Together, the top three DR methods per field are cited more than the bottom 75 methods in \bio, Medicine, Psychology, Agricultural and Food Sciences, Linguistics, Sociology, Geography, and Law.
This may suggest that in these subject areas, only a small set of DR methods are established or accepted or that data characteristics in such areas are amenable to certain DR methods.

We next consider how the 78 DR methods have been cited over time.
We use percentile rankings of citation counts stratified by subject area and year to enable temporal comparison. 
Focusing on t-SNE, PCA, and UMAP in \bio, \phys, \busi, and \chem, we find that percentile rankings for t-SNE, PCA, and UMAP increased between 2013 and 2022 in \bio, \phys, and \chem. In \busi, percentile rankings for t-SNE and UMAP increased while those for PCA stayed consistent (\autoref{fig:bibplot_time}).
These results are consistent with the publication of t-SNE in 2008 and UMAP being preprinted in 2018.
The general trend between 2013 and 2022 across subject areas is that percentile rankings of DR methods have increased over time.

In summary, we find that while the use of \dr is prevalent and generally rising across most fields (\rqone), the techniques used vary greatly by field (\rqthree).  In fields relating to biology and medicine, t-SNE and UMAP are popular techniques, while in fields relating to physics, chemistry, or business, PCA is popular.  While there are other techniques used, t-SNE, PCA, and UMAP are the three most cited methods in recent years.  



\section{Subject Area Survey}
\label{sec:full-analysis}

The bibliometric analysis provided some insights into the broad trends and practices across domains outside of computer science \paran{\rqone} but did not provide insight into the types of data used or nuances about the usage of particular techniques.  Likewise, such a high-level analysis could not analyze the visual interpretations offered \paran{\rqtwo}.  As a subsequent study, we conduct a close reading of a survey of literature in research outside of computer science.

\subsection{Grounded Analysis}

We began with a grounded analysis~\cite{starks2007choose, diehl2022characterizing} to consider a theory for analyzing and comparing works based on their \drshort.  Our goal with this initial investigation was to ground the analysis of our subsequent literature review, bootstrapping the dimensions on which the usage and interpretation of dimensionality reduction techniques were spread among different scientific disciplines.  

For this grounded analysis, we loosely gathered papers from varied domains using Google Scholar and Scopus.  In both search engines, we searched for articles listing the different \dr techniques in their title, abstract, or keywords and with subject areas of Arts and Humanities, Economics, Econometrics and Finance, Nursing, Physics and Astronomy to get a varied collection of papers.  Between these two searches, 
we found articles in single-cell genomics~\cite{becht2019dimensionality, ujas2023guide, cieslak2020t, clarke2021tutorial, andrews2021tutorial, kimball2018beginner, liechti2021updated, dorrity2020dimensionality}, business/finance~\cite{lee2023can, vriens2019mapping}, urban studies~\cite{wang2022unsupervised}, physics~\cite{ch2018unsupervised, traven2017galah}, and epidemiology~\cite{sakaue2020dimensionality}.  

In our reading of these papers, we found that the usage of dimensionality reduction techniques varied in the analytical goal, the techniques used, the preprocessing and parameterization of those techniques, and the interpretation of the results of those techniques.  We also found some similarities and differences between the scientific subject areas.  For example, many single-cell genomics papers used a 2-D scatter plot of dimensionally reduced data as one step in a larger analytics pipeline, only gathering hypotheses to later test, while other domains drew conclusions directly from the reduced space.  Similarly, the types of data, including the number of rows and columns, seemed to naturally lead to different types of interpretations.  The design decisions of the visualizations used also seemed to vary, as some papers would rely on many annotations and highlights to explain how the dimensionally reduced data should be interpreted, while others provided only a few sentences in the main text.  In particular, we found that the usage of \drshort differed primarily in the \textbf{data} shape and \drshort algorithm being used on that data, the \textbf{design} of the visualization of the dimensionally-reduced data, the low-level \textbf{tasks} used in the interpretation of that visualization, and the larger \textbf{workflows} that the \drshort was a part of.  

\subsection{Selection of papers}

\label{sec:selection}

Drawing from our bibliometric analysis, we first decide on four subject areas to focus on based on their diversity in \drshort usage: \bio, \chem, \phys, and \busi.  In particular, we found that \bio was the field that cited the top three methods the most and featured a large body of research, while \chem, \phys, and \busi were fields that featured usage of alternative \drshort algorithms.  In addition, several of the coauthors of this work had experience working with domain scientists in these four fields.  
While there were other fields that were more popular or featured unique usage, such as Environmental Science, they were either very similar to the chosen fields or they were outside of the area of expertise of the co-authors.  We note that this selection of four domains does limit the generalizability of our results to just those domains.

Then, we use a literature database to search for papers matching keywords relevant to dimensionality reduction.  We required that any paper in our analysis i) uses a dimensionality reduction algorithm as part of its data analysis and ii) presents a visualization of the dimensionally reduced data.

We use the Preferred Reporting Items for Systematic Reviews and Meta-Analyses (PRISMA) guideline to conduct systematic reviews~\cite{page2021prisma}.  We chose Scopus\footnote{Query strings included in the appendix.} as our literature database, based on its reproducibility and its wide coverage of the subject areas we chose~\cite{gusenbauer2020academic}. The search strategy searches for keywords ("Dimensionality Reduction", "UMAP," "t-SNE," "PCA," "Projection") within Scopus subject areas of chemistry, biology, physics, and business within the last five years.  Subject areas were based on the Scopus database labels.  These projection techniques were used based on the bibliometric findings that these were the most common methods.  In addition, these methods have been studied previously by the visualization community.

Our initial Scopus search returned 52141 matching papers.  We then filtered to only those papers with more than 5 citations, according to Scopus, to filter out papers with low impact, reducing the number of publications to 21083.  Next, we conducted a stratified sample from this set down to 2000 publications, with 500 from each of our 4 subject areas.  From this set of 2000 publications, we used the Zotero Reference Manager's\footnote{https://www.zotero.org/about/} \textit{Find Available PDF} function from our academic library's connection to identify 930 of those publications for which we could easily find the PDF for review.  While the additional PDFs could eventually be found, we believe that this filtering to 930 publications should serve as a fairly uniform subsampling of the 2000 publications, which was appropriate since we ultimately sought to select just a small sample of the papers.  This sampling is allowable because we were not aiming to completely survey all works using \dr in these fields; rather, we aimed to merely sample them.

We divided these 930 publications amongst the four authors and scanned them to filter out any publications that did not have any visualization of the dimensionality reduction results, which removed 62\% of the publications.  Then, from the resulting set of 347 publications, each author randomly selected five publications from each of the four subject areas, as designated by Scopus.  This resulted in 71 papers being reviewed: 20 \bio, 20 \chem, 17 \busi, and 14 \phys~ (upon closer reading, 3 out of the 20 \busi~ and 6 out of 20 \phys~ were incorrectly classified by Scopus and instead came from other fields, i.e. industrial design and engineering, and so were excluded from our analysis).  

We discuss insights from this paper's winnowing process in section~\ref{sec:discussion}. We note that the relatively small sample of 71 papers included in our survey are not completely representative of the more than fifty thousand papers returned in our initial search in Scopus. However, we also believe that 71 papers is a large enough sample to provide valuable insights into the usage of \drshort outside of computer science.  In addition, we investigate the author list of these 71 papers to ensure that there is not an overrepresentation of any particular authors.  Across these 71 papers, there were 705 unique authors with only a single author appearing on more than one paper (two).  We believe this sample represented a diversity of authors across these domains.  In addition, we analyzed the listed affiliations of these papers and found that only 6 out of 705 unique authors listed an affiliation in a department of computer science or information science \cite{cheng2018monitoring, kuchroo2022multiscale, lee2019dynamic, wei2021novel, xu2022pattern, xu2020t}.

\section{Findings}
\label{sec:findings}

In this section, we describe the findings from our literature review.  We organize our findings based on our grounded analysis found in \autoref{sec:full-analysis}.  We present findings related to the \textbf{data} being projected, the \textbf{design} of the visualization of the projected data, the \textbf{tasks} used in interpreting the projection according to the in-line text and captions, and the high-level \textbf{workflows} that the projections serve within the flow of the publications.  We list the subject area when citing works from the survey to identify similarities and differences (\textbf{RQ3}) observed during our in-depth review.  Classification results can also be viewed and explored in our online browser, available at \url{https://dimension-reduction-vis.github.io/}.  This browser includes screenshots of visualizations from each paper that we read as part of this report.

\subsection{Data}

\subsubsection{Data Shape}

We recorded the number of rows (i.e., points) and columns (i.e., dimensions) of the data fed into the \drshort method if they were reported by the authors.  The data varied significantly across the 71 publications we surveyed.  The size of the data ranged from as few as 7 data points~\cite{babinszki2020comparison} to more than twenty million~\cite{kuchroo2022multiscale}.  The number of dimensions of data ranged from 4~\cite{lin2020exploring, malafronte2021integrated} to more than 13 thousand~\cite{potluri2020antibody}.  

\begin{figure}
    \centering
    \includegraphics[width=\linewidth]{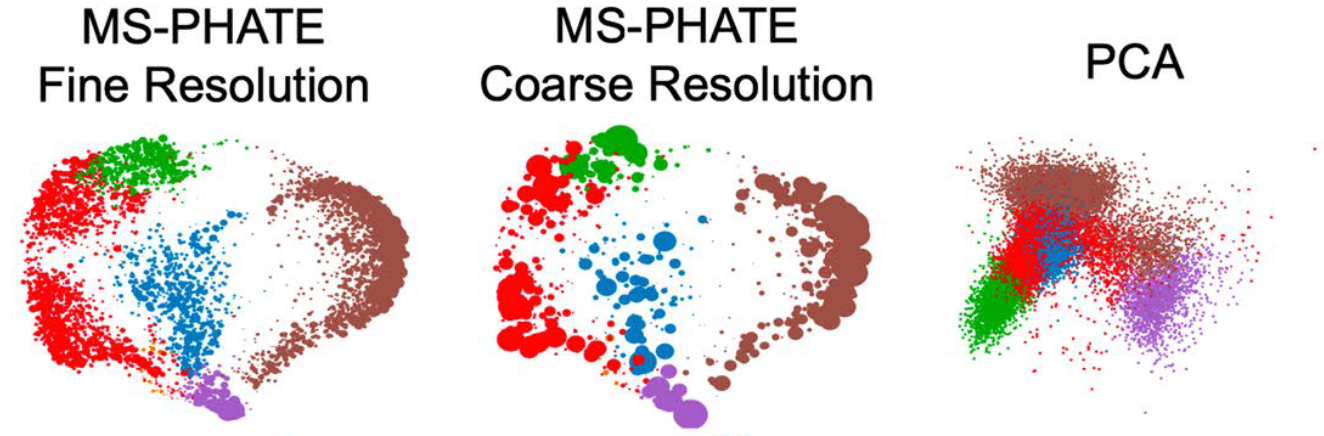}
    \caption{Three plots of 22 million measurements of peripheral blood mononuclear cells, or PBMCs, gathered via flow cytometry~\cite{kuchroo2022multiscale}.  When the number of elements in the projections is high, transparency is often used so that the general distribution of the data in the projected space can be seen.}
    \label{fig:many_points}
\end{figure}

When there are many data points, points are drawn with some level of opacity to communicate the density of the projected space as in \phys Cheng et al.~\cite{cheng2018monitoring}, as in  \autoref{fig:many_points}.  In studies with few data points, the data points were typically categorized to understand which points were similar and which were different.  For example, in Hasan et al.~\cite{hasan2020accumulation} (\phys), eight different metals were analyzed by their concentrations found in either soil samples (SS) or food samples (FS) grown in that soil.  PCA is used to project those samples into a two-dimensional space, and the relative location of the two types of samples in the projected space is used to conclude that the distribution of metals was different.  
Similar phenomenon was observed also for \chem (e.g., relating 8 different states of a kinase~\cite{wang2019globally}), \busi (e.g., analyzing relationships between the economies of European nations~\cite{jonek2022assessing} or Chinese corporate brands~\cite{lin2020exploring}), and \bio (comparing different strains of bacteria~\cite{jin2021evaluation}).

\subsubsection{\drshort Method}

We also examined the type of \drshort method used to process the data.  We identified that higher dimensional data necessitates domain-specific and/or nonlinear \dr techniques and that the types of data found in different subject areas drove different methods.  \bio and \chem papers tended to employ more nonlinear \dr methods (e.g., UMAP, t-SNE).  The \bio~publications that reported input dimensions reported an average of $\mu=2186$ dimensions and \chem~$\mu=751$, compared to \busi~$\mu=19$ and \phys~$\mu=12$.  It is likely that the phenomena being captured in high dimensional data are not often visible or might be difficult to identify in a linear projection like PCA. 
For example, in the \chem field, Mazher et al.~\cite{mazher2020visualization} analyzed a dataset of 5,688 data points with 273 dimensions, comparing different non-linear dimension reduction methods, including UMAP, t-SNE, and a newer technique, Potential of Heat-diffusion for Affinity-based Trajectory Embedding (PHATE).  The choice of the \drshort technique may also be the result of disparate tasks in each subject area, which we discuss shortly.

Surprisingly, in each domain, there were examples of papers that used multiple dimensionality reduction techniques because the techniques were seen as being useful for different types of data.  An example is found in \phys~Lee et al., in which various data extracted from a small sample of newborn blood is analyzed~\cite{lee2019dynamic}.  The blood samples were processed to read several categories of data: (1) cellular composition, (2) plasma cytokines/chemokine concentration, and (3) several other biological measures like protein composition and metabolomic data.  In this case, PCA was used to demonstrate the separability of the data for categories (1) and (2).  However, the authors suggested that a domain-specific technique, Data Integration Analysis for Biomarker discovery using Latent cOmponents (DIABLO), designed for biomarkers, was needed when integrating the third category of data because of \textit{"the complexity of the data \ldots and the heterogeneous nature of data measured on different scales and technological platforms"}. 

In 14/71 (20\%) of the papers in our survey, a technique besides PCA, tSNE, or UMAP was used.  This alternate technique was frequently a factor analysis, which is a statistical technique that uncovers primary factors that result in the separation of data points.  In this analysis, the dimensions of the data are combined in a linear combination into two different components similar to PCA, but use a different method that is more standard within their domain, such as orthogonal projections to latent structures discriminant analysis (OPLS-DA) within~\bio~\cite{carbajo2019beneficial}.  Single-cell \bio is unique in that some nonlinear techniques have been developed by computational biologists that are specially designed for high-dimensional biology data~\cite{moon2019visualizing, kuchroo2022multiscale}.


\subsection{Design}

We categorize the design of the visualization of the \drshort view by its plot type, its plot style, and the annotations used.

\subsubsection{Plot Type}

\label{sec:plottype}


\begin{figure*}
    \centering
    \includegraphics[width=\linewidth]{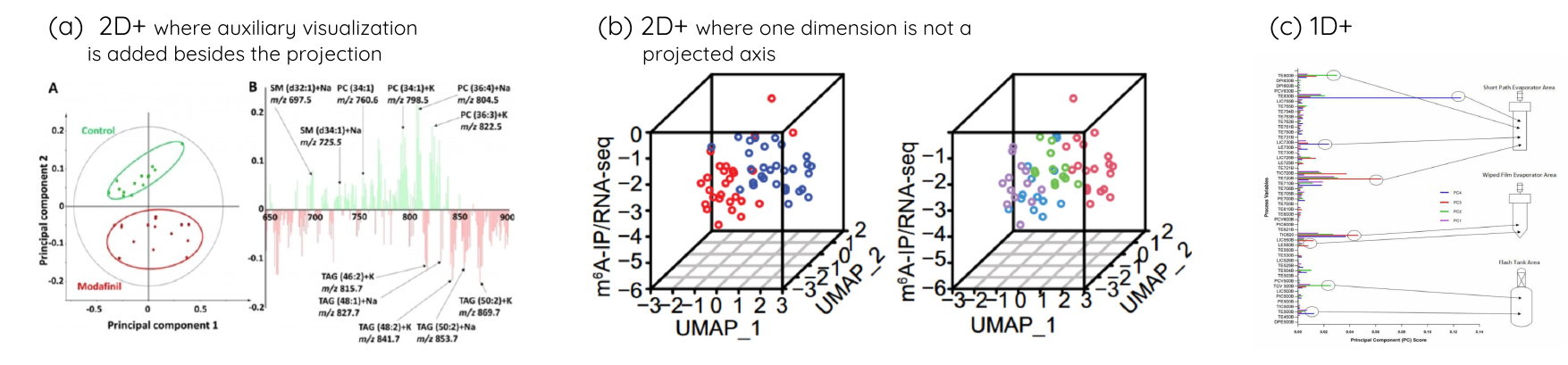}
    \caption{The example figures for 2D+ and 1D+ plot types (\autoref{sec:plottype}). (a) 2D+ plot in which an auxiliary \textit{plot }is added to augment the projection. (b) 2D+ plot in which an auxiliary \textit{axes} is added to represent the dimensions that are not a projected axis. (c) 1D+ plot consists of a 1D projection and an auxiliary plot.  Found in ~\cite{philipsen2021mass, yao2023scm6aseq, teng2019principal}, respectively }
    \label{fig:plottypes}
\end{figure*}

  We consider four plot types, all variations on the scatter plot.  3D and 2D describe scatterplots with three and two dimensions, respectively.  2D+ describes a two-dimensional scatterplot integrated with an additional plot, as in \autoref{fig:plottypes}a~\cite{wauters2021discriminating} (a), or if a three-dimensional plot is used where one of the dimensions is not a projected axis, but instead used to view correlation as in~\chem~\autoref{fig:plottypes}b~\cite{yao2023scm6aseq}.    Likewise, 1D+ features a one-dimensional plot of the data points, as seen in \busi~\autoref{fig:plottypes}c~\cite{otten2021event}.  The overwhelming majority (63/71) of papers included a 2D plot, with 9 3D plots, 8 2D+ plots, and 7 1D+ plots.  

\subsubsection{Plot Style}

\begin{figure*}[ht]
    \centering
    \includegraphics[width=\linewidth]{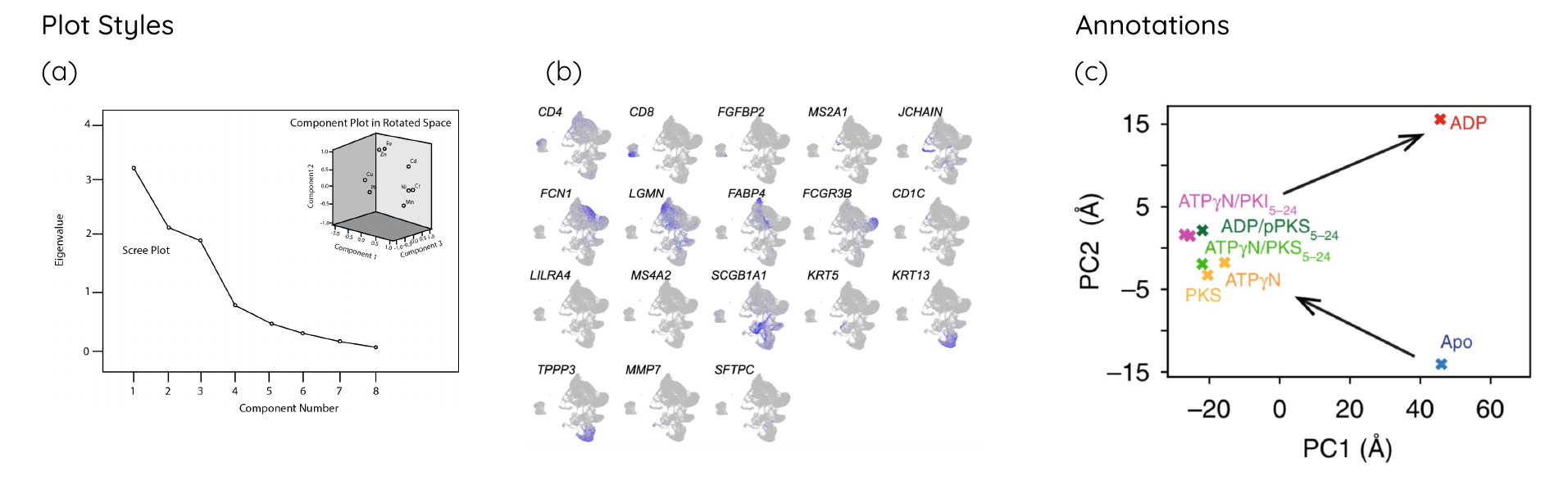}
    \caption{The example figures for different plot styles and annotations. (a) Small multiples leveraged to analyze more than two composite dimensions~\cite{hasan2020accumulation} (figure manually reproduced by authors due to copyright claims) (b) Small multiples to compare different variables at once~\cite{wauters2021discriminating} (c) Annotations for depicting the connections between data items~\cite{wang2019globally} }
    \label{fig:plotstyle}
\end{figure*}

We report whether plots have a legend, drawn axes, gridlines, and small multiples of scatterplots.  Legends were less common in \busi~(24\%) and \chem~(50\%) than \phys~(57\%) and \bio~(85\%).  Axes were usually drawn (89\%), while gridlines were less frequent (37\%), although they were markedly more frequent in \busi~(65\%).  Small multiples were rarely used in \busi~(12\%), \chem~(25\%), and \phys~(36\%), but frequently in \bio~(65\%).  In some cases, small multiples were used to analyze more than two composite dimensions of the \drdata as in~\busi~\cite{yang2020power} or~\phys~\cite{hasan2020accumulation} (\autoref{fig:plotstyle}a), while in other cases, the same view of the \drshort space is presented multiple times with different variables encoded by color as in~\bio~(\autoref{fig:plotstyle}b)~\cite{wauters2021discriminating}.

We additionally record whether the plots are annotated using glyphs or symbols to signify another variable, textual labels directly on the plot, highlighting, and captions.  Captions, or textual descriptions spatially attached to the figure, were frequently used across our survey corpus (82\%).  Textual labels directly annotating visual elements (51\%), highlighting (37\%), and glyphs or symbols (24\%) were used less frequently with no clear patterns of difference between different subject areas.  Annotations were commonly used to highlight clusters and identify relationships between clusters and other variables.  In particular, many papers including~\chem~\cite{collins2020comparison} and~\bio~\cite{bernardo2019untargeted} would draw enclosing circles around clusters or to signify 95\% confidence ellipses, and also use color and shape to show additional attributes beyond those used in the projection.  These annotations were often explicit, using text directly on the plot to label clusters.  In some cases, arrows or connecting edges were drawn to show connections between data in two different known groups, as seen in \autoref{fig:plotstyle}c.  Axis titles often note the percentage of variance explained using text. We observe that authors also draw quadrants and annotate different sections with interpretation. There is no one common way of explaining axes.



\subsection{Tasks}

\subsubsection{Task Descriptions}

\newcommand{\taskicon}[1]{%
  \begingroup\normalfont
  \includegraphics[height=\fontcharht\font`\B]{#1}%
  \endgroup
}

Through our grounded analysis and our close reading of our survey of subject areas,  we identified seven common tasks that were commonly used in describing how the plots of \drshort data should be interpreted.  Each task is illustrated and described below, and the percent of papers the task is found in is reported.  To address \rqfour, we compare our seven tasks with the tasks identified in three previous visualization works: Etemadpour et al.~\cite{etemadpour15tvcg}, Nonato \& Aupetit~\cite{nonato18multidimensional}, and Xia et al.~\cite{xia22tvcg}.  Notably,  we only consider tasks that are relevant to the interpretation of static views of \drdata, rather than those tasks available in interactive systems.  As a result, we exclude tasks that explicitly refer to user interactions with \drshort, such as those from Sacha et al.~\cite{sacha2016visual}.

\vspace{3mm}

\begin{wrapfigure}{L}{1.2cm}
    \vspace{-\baselineskip}
\includegraphics[width=1.2cm]{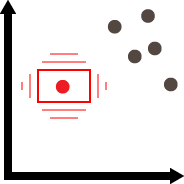}
\end{wrapfigure}
\textbf{Single Point} (9/71 = 13\%).  In this task, a single data point is highlighted and described, whether in the main text or within a caption.  Its place within the projected space is used to provide insight into the particular point, as in \phys~identifying the distance of a control point from the rest of volatile compounds in the \drshort view~\cite{nunezcarmona2019innovative}.  The closest proposed task is from Etemadpour et al.~\cite{etemadpour15tvcg} to identify the closest cluster to a given object (\textit{fCluObj}). Nonato \& Aupetit mention a similar task of \textit{finding a seed point}, and Xia et al. do not include a task on identifying a single point. We find that the interpretation of a single point is not a common task within our survey, but when it is a task, it typically involves identifying proximity to a cluster.

\vspace{3mm}

\begin{wrapfigure}{R}{1.2cm}
    \vspace{-\baselineskip}
\includegraphics[width=1.2cm]{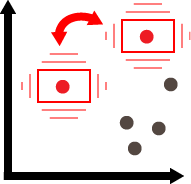}
\end{wrapfigure}
\textbf{Multiple Points} (8/71 = 11\%).  In this task, multiple data points are described and compared.  This comprises the tasks of identifying nearest neighbors from Etemadpour et al.. but can also be broader to account for expected relationships between points.  For example, in \bio~Cui et al., particular macrophages are identified in a PCA view as being related to pulmonary fibrosis lungs~\cite{cui2020activation}.  In \busi~Feuillet et al., ten years of French soccer club seasons are projected into a PCA plot, and multiple points corresponding to multiple seasons of the same club are analyzed to understand changes in strategy over time~\cite{feuillet2021determinants}.  This example may be similar to the task of Nonato \& Aupetit of \textit{identifying a path} within \drdata.  Again, this is not a common task, suggesting that individual points are not typically the object of analysis within our survey.

\vspace{3mm}

\begin{wrapfigure}{L}{1.2cm}
    \vspace{-\baselineskip}
\includegraphics[width=1.2cm]{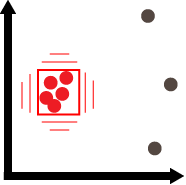}
\end{wrapfigure}
\textbf{Single Cluster} (23/71 = 32\%).  In this task, multiple data points are described and compared based on their visual clustering within the projected view.  It is identified by both Nonato \& Aupetit and Xia et al. as an exploration task to discover a cluster within a projected view.  In Etemadpour, this task is separated into two subtasks: \textit{\#SClu} i.e., estimating the number of subclusters within a cluster, and \textit{\#Obj} estimating the number of objects within a cluster.  However, we did not find counting to be a common action in analyzing a cluster.  Instead, a cluster might be analyzed to develop an explanation for the isolation of some particular subgroup within the data.  Examples include~\bio~Potluri et al. identifying a set of patients with a particular metastatic disease phase being visually clustered together in a PCA view of antibody profile data~\cite{potluri2020antibody}, or~\chem~Wang et al. identifying a cluster of transitional states of an enzyme as shown in \autoref{fig:plotstyle}c~\cite{wang2019globally}.  

\vspace{3mm}

\begin{wrapfigure}{R}{1.2cm}
    \vspace{-\baselineskip}
\includegraphics[width=1.2cm]{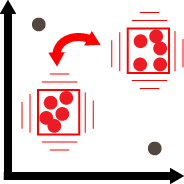}
\end{wrapfigure}
\textbf{Multiple Clusters} (46/71 = 65\%).  In this task, multiple clusters are analyzed.  This could be to compare different groups.  This is found in \chem comparing treated vs. untreated samples~\cite{babinszki2020comparison,yuan2020modification} and different organic sample locations~\cite{castro2020metabolomics, duan2021genotypic}.  It is also found in~\phys to compare different types of celestial bodies~\cite{clarke2020identifying}.  Alternatively, it could be to try to define the different clusters that emerge.  As an example, in \busi~\cite{ji2021insights}, Ji et al. interpreted the distance between clusters of different types of biofuel as being greater than the distance between clusters based on the temperatures at which those fuels were burned.  This was the most common task found in our survey, but the comparison of clusters is not explicitly included as a task in any of the visualization works we compared to.  Closest is the \textit{distance comparison} from Xia et al., but that task specifically refers to identifying the closest cluster to a given cluster rather than interpreting the distance and relative locations of two arbitrary clusters.  This may not be a focus of visualization research because it is fraught with potential misinterpretations we describe in \autoref{sec:incorrect_usages}, including the interpretation of global distances and nonlinear axes.

\vspace{3mm}

\begin{wrapfigure}{L}{1.2cm}
    \vspace{-\baselineskip}
\includegraphics[width=1.2cm]{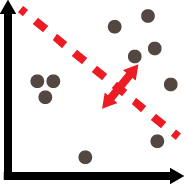}
\end{wrapfigure}
\textbf{Global Patterns} (34/71 = 48\%).  This task was the second-most-common of the seven we identified.  In this task, different regions of the projected view independent of particular clusters are ascribed meaning.  This typically involved interpreting the axes or quadrants of the projected view in order to explain clusters or verify known relationships in the data.  Related subtasks for interpreting global patterns are identified in Nonato \& Aupetit including \textit{naming} and \textit{discovering relationships} between reduced dimensions and original dimensions.  Examples can be found in \chem in interpreting different clusters of states of matter~\cite{kriston2020analysis} or locations where samples were taken~\cite{ma2019analysis} and likewise in \phys to explain principal components of climate data~\cite{xu2019synchronous}.

\begin{figure}[ht]
\centering
\includegraphics[width=\columnwidth]{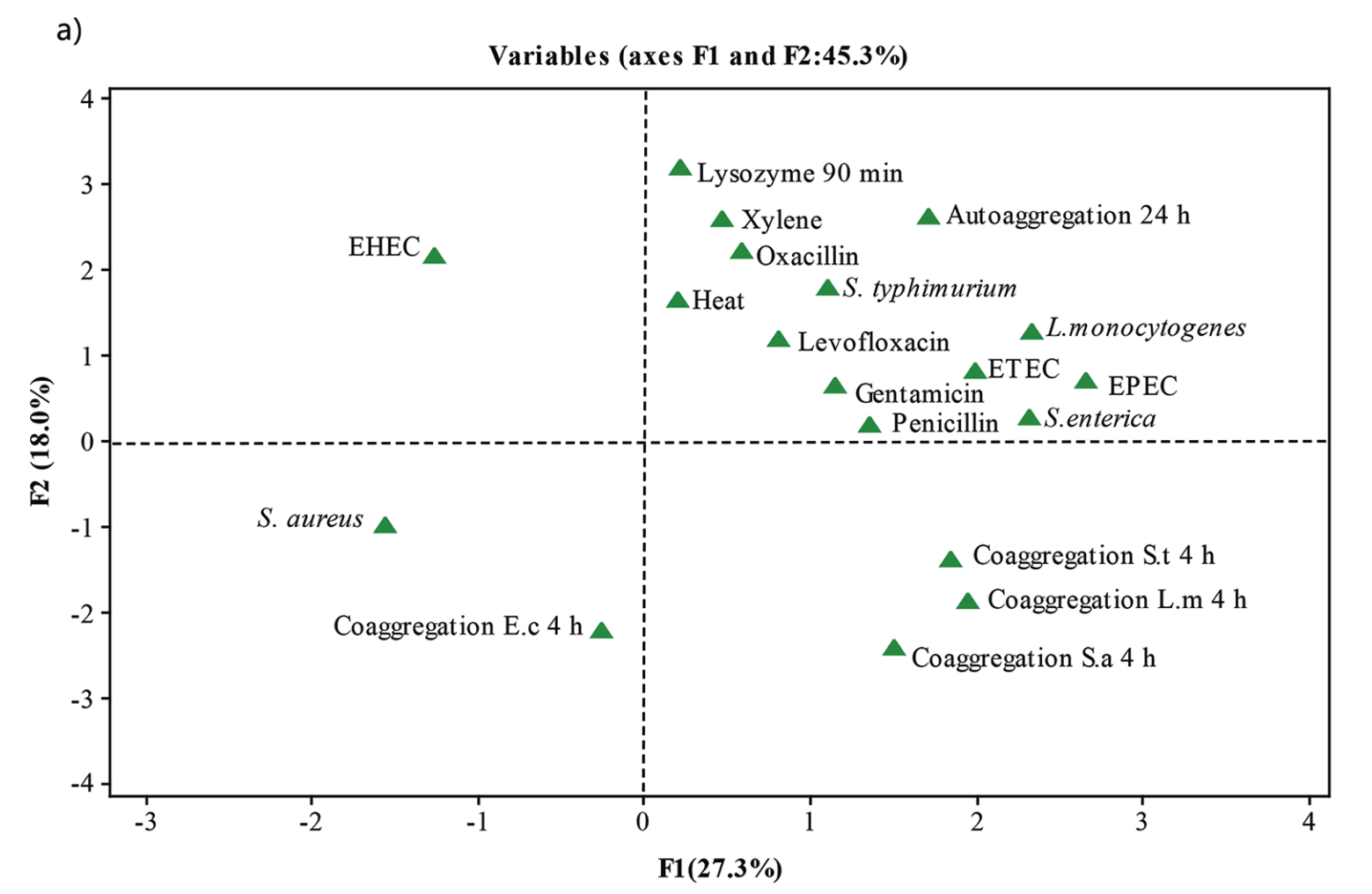}
    \caption{In Jin et al.~\cite{jin2021evaluation}, quadrants of a PCA loading plot (in which points represent features) are used to understand how features are (anti-)correlated with the first two principal components.}
    \label{fig:jin2021evaluation}
\end{figure}

\begin{figure}[ht]
\centering
\includegraphics[width=\columnwidth]{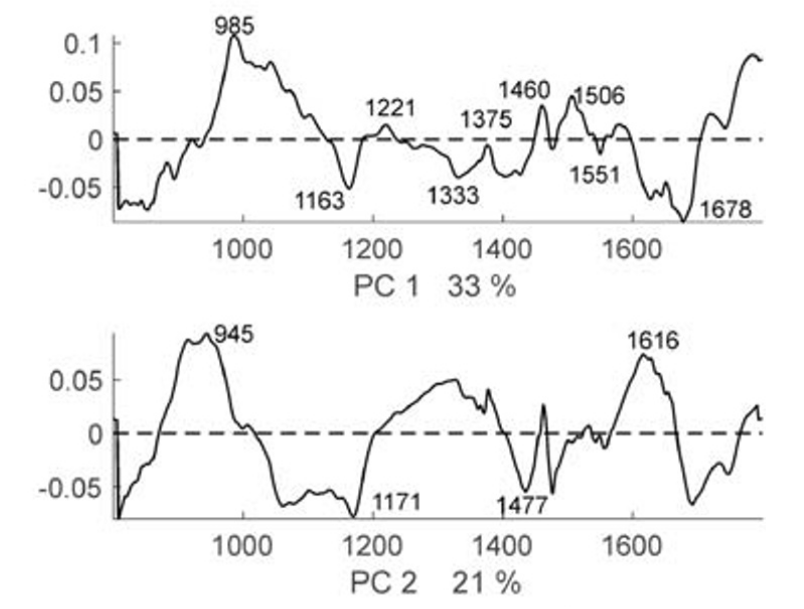}
    \caption{In Diehn et al. \cite{diehn2020discrimination}, spectra from pollen samples comprising five species of grass were obtained using Fourier-transform infrared spectroscopy (FTIR) for chemical characterization. PCA was applied to a dataset of fifty observations (each observation representing the average of 20 pollen grains from a single plant) of the spectral range from 800 to 1,800 cm$^{-1}$. The authors plot the loadings for PCs 1 and 2 against the measured spectral range.  The authors note that two groups of grass species (which appear in different regions of a PC1-PC2 scatterplot) can be distinguished from each other by taking advantage of the extreme values observed in the loading plots around 1,678 cm$^{-1}$ for PC1 and around 945 cm$^{-1}$ for PC2, which correspond to known molecular vibrations of proteins and carbohydrates.}
    \label{fig:diehn2020discrimination}
\end{figure}

In some cases, more than two dimensions of the \drdata are interpreted.  \bio~Jin et al. \cite{jin2021evaluation} analyze the first four principal components for correlations with input features using a loading plot, seen in \autoref{fig:jin2021evaluation}, which then informs the interpretation of the corresponding PCA view.  Similar analyses of the principal components are split into one-dimensional plots shown in \autoref{fig:diehn2020discrimination} side-by-side with traditional two-dimensional scatterplots (not shown) in \bio~\cite{diehn2020discrimination}.

The analysis is not only axis-aligned, as in \chem~Wang et al.\cite{wang2019globally} (\autoref{fig:plotstyle}c), where diagonal directions are interpreted as state transitions of an enzyme.  This type of linear interpretation of a space is similar to the identified task of \textit{discovering a path} within the projected view from Nonato \& Aupetit.

We believe that the popularity of this task within our survey points to unclear guidelines on how to interpret \drdata.  The potential distortions outlined in Nonato \& Aupetit indicate that the broad analysis of the \drdata can result in misconceptions about the source data.  We believe that there are opportunities for visualization researchers to provide more precise interpretations via the mitigations suggested in Nonato \& Aupetit, which we describe in \autoref{sec:discussion}.




\vspace{3mm}

\begin{wrapfigure}{R}{1.2cm}
    \vspace{-\baselineskip}
\includegraphics[width=1.2cm]{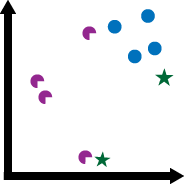}
\end{wrapfigure}
\textbf{Relationship to Other Variables} (29/71=41\%).  In this very common task, points in the projected view are additionally encoded with some additional attribute and the projected view is compared to any natural ordering provided by the additional encoding, commonly to verify that the projection separates the data by that additionally encoded data.  For example, in~\bio~Ocasio et al. (\autoref{fig:incorrect_usages}, far right), scatterplot points are colored by expressed transition state, and the transition of colors in the plot is annotated to demonstrate that the projected view captures the phenomenon of state transition.  Often, a categorical class is used in a symbol or color encoding, which was found in \busi~\cite{nam2019global}, \bio~\cite{du2021analysis}, and many others.  In other works in \bio, it was common to present complex data in small multiples with each scatterplot encoding a different variable, as seen in \autoref{fig:plotstyle}b~\cite{wauters2021discriminating} and others~\cite{manco2021clump, aissa2021single, sengupta2022mesenchymal}.  This technique was sometimes found in \chem as well~\cite{moosavi2020understanding}.  This very common task was not included explicitly in the visualization works we compared to, and we believe it should be an opportunity for research, as the use of glyphs, symbols, and other encodings can affect the perception of \drdata.

\vspace{3mm}

\begin{wrapfigure}{L}{1.2cm}
    \vspace{-\baselineskip}
\includegraphics[width=1.2cm]{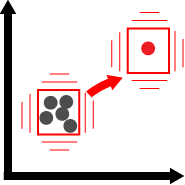}
\end{wrapfigure}
\textbf{Outliers} (5/71 = 7\%).  In this last task, outlier points in the projected view are described to interpret the data projection.  In \busi~Onuferová et al., Slovakian travel agencies are analyzed both statistically and visually.  Visual outliers within a PCA view are then termed as "extremes"~\cite{onuferova2020analysis}.  In a similar manner, \busi~\cite{vstefko2021application} highlight outliers in a PCA view of a data envelopment analysis of heat management companies to identify companies that are at risk of bankruptcy.   The identification of outliers is a task described by Nonato \& Aupetit.  The interpretation of outliers is related to tasks identified by Etemadpour et al. (\textit{fCluClu}) and Xia at el. (\textit{membership identification} and \textit{distance comparison}) in which a point is identified as being close to one cluster vs. another.  However, in our survey we found that points far from clusters (i.e. outliers) were more frequently described rather than those close to cluster centers.

\subsubsection{Summary of Differences from Prior Works}

Compared with Etemadpour et al., we did not find that the count of the number of subclusters (\textit{\#SClu}), the number of outliers (\textit{\#Out}), the number of objects within a cluster (\textit{\#Obj}), or the relative densities of clusters (\textit{rDens}, also identified as a primary task in Xia et al~\cite{xia22tvcg}) to be commonly analyzed~\cite{etemadpour15tvcg}.  In contrast, in our survey we found that the analysis was generally coarse, describing a phenomenon found across a dataset, and so the count of objects within clusters (or their visual density) wasn't discussed.  In addition, outliers were identified in smaller datasets in several \busi works, but they were analyzed individually rather than counted.  

Besides \textit{density comparison}, Xia et al noted three additional typical tasks: \textit{cluster identification}, \textit{membership identification}, and \textit{distance comparison}.  The identification of clusters is found in our literature review, but the measurement of distance and the cluster membership of individual points are not commonly found.   

Nonato and Aupetit present a much more complete list of 32 tasks, with one subgroup group of 8 tasks, \textit{Explore Items in Base Layout}.  These tasks include \textit{Discover Clusters}, \textit{Discover Paths}, and \textit{Discover outliers}.  However, it also includes interactions such as navigation and brushing, and some more local investigation into neighborhoods.  In addition, out of the more than 40 papers surveyed, the tasks were only found in a maximum of 7 papers in their survey, suggesting that they were not prevalent tasks in the visualization literature.

We believe that the analyses and interpretations we observed in our literature review represented a different sample than those surveyed in the four prior works, which observed the use of \dr in visualization research.  In our observations of our literature review, counts of clusters or objects may not have been important because they are not useful in confirming prior hypotheses about the data.  The authors typically used the presence of clusters or the separability of clusters to provide evidence that particular phenomena well-known in their communities (like the difference between cell types or physical sample sources) are identifiable in the data.  The number of individual subclusters is not relevant to these types of hypotheses.  The place of \dr in workflows within visualization papers may be different than the place found within domain papers.

The four prior works are not a complete union of the proposed task analyses of \dr use cases, although we note that Nonato and Aupetit do cite many design studies in their analysis.  It is likely that there have been design studies, including collaborations between visualization researchers and domain scientists.  We do not include individual design studies in the scope of this paper.  However, future work in surveying design studies in visualization research could potentially surface novel tasks.  While there are existing surveys on \dr in visualization research, they do not include a task meta-analysis.  In section~\ref{sec:discussion}, we further discuss the role of design studies and surveys in our opportunities for visualization researchers.




\begin{figure*}[t]
    \centering
    \subfloat[]{
        \includegraphics[height=2.4in]{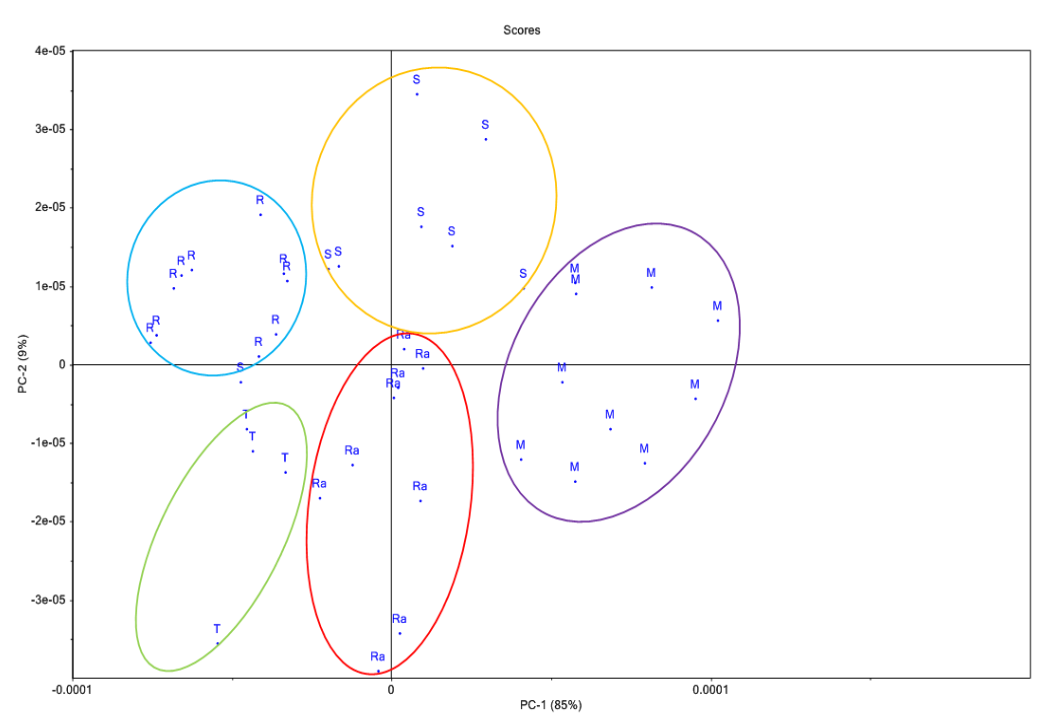}
    }\hskip4ex%
    \subfloat[]{
        \includegraphics[height=2.4in]{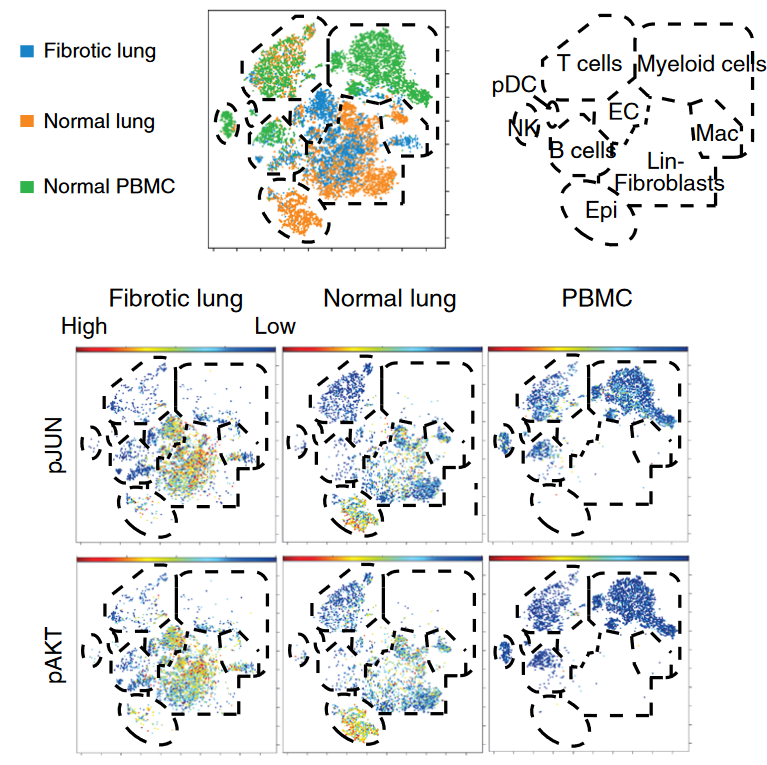}
    }\hskip4ex%

    \caption{We identify three common workflows for using visualization of \dr results, two of which are shown here and one is evident in a previously highlighted figure in \autoref{fig:plottypes}a.  The three workflows differ based on whether the visualization is used for exploratory data analysis as in \autoref{fig:plottypes}a~\cite{philipsen2021mass} where the relationship between frequencies in a spectroscopy and clusters is being explored, confirmatory data analysis as in (a) of this figure ~\cite{pauliuc2020raspberry} where known groupings in the data are verified to exist in the projected space, or a combination of both as in (b) of this figure~\cite{cui2020activation}.  In this combined workflow, known groupings are first used to informally validate the layout.  Then, the relationship between additional variables (in this case, the activations of two transcription factors in different types of tissue) is explored in the projected space.}
    \label{fig:workflows}
\end{figure*}

\subsection{Workflows}

Across domains, we encountered three common workflows where dimensionality reduction results were explicitly visualized in describing a data analysis process.  These contrast in some ways to popular understandings of the role of visualization, such as Pirolli and Card's sensemaking loop~\cite{pirolli2005sensemaking} or Van Wijk's model of the value of visualization~\cite{van2005value}.  These models could suggest that the visualization of high-dimensional data in a 2D space would largely be done for exploratory data analysis, to make sense of the data, understand trends, and generate hypotheses.  However, in our analysis, we found that there was often a mix of confirmatory data analysis, in which the generated projection was used as an ad-hoc statistical test~\cite{wickham2010graphical}.  In this case, the visual separation of clusters, for example, could be used as evidence that data is separable in the high-dimensional space.  This separability is then extrapolated to make a judgment about the value of the data for later analysis.

\noindent \textbf{Workflow 1: Exploratory Data Analysis} One common workflow is to use a projection for exploratory data analysis to generate hypotheses, which are then verified by other statistical tests.  In this workflow, a complex phenomenon is being studied to develop a greater understanding, and the goals can be developed iteratively as the data is explored.  The workflow is typically seen when a rich dataset was assumed to be related to a particular phenomenon, but the causality or model connecting the data to the phenomenon was not known, as in \bio~mass spectrometry data~\cite{philipsen2021mass} and gene expression data~\cite{munoz2019coordinated}, \phys~remote-sensing reflectance data~\cite{lange2020radiometric}, or heuristically gathered data in \busi such as forest management~\cite{riccioli2020indicators}, agricultural~\cite{berton2021environmental} or nutritional~\cite{moreira2019evaluation} indicators, and survey data~\cite{maghlaperidze2021development, hetenyi2019quantitative}.  

For example, in Philipsen et al.~\cite{philipsen2021mass}, mass spectrometry imaging data of the fly brain is used to understand the effect of a treatment, modanifil.  This generates hundreds of spectra, each one a dimension of data, and the goal of the analysis is to understand which spectra might change in response to the treatment.  As seen in Figure~\ref{fig:plottypes}a, the spectroscopy data is projected down to two dimensions with PCA, showing separability in the second principal component.  Then, a loading plot shows which spectra are most discriminatory in the second principal component.  This generates hypotheses that the bands of spectra that peak or valley may quantify the effect of the treatment.  After identifying these spectra, the authors explain "\textit{it is hard to interpret the precise differences based on the scores and loadings.  Hence, the changes in the level of each molecular species are measured and evaluated with statistical analysis.}"

Similar workflows were used in \phys literature.  As described in Conterosito et al.~\cite{conterosito2020situ}, \dr can be applied to physical data to "speed up analysis with the specific goals of assessing data quality, identifying patterns where a reaction occurs, and extracting the kinetics."  This type of analysis was also used to generate hypotheses in Kobaka et al.~\cite{kobaka2021principal} to identify differences in concrete mix designs.

\noindent \textbf{Workflow 2: Confirmatory Data Analysis}  In the second type of workflow, the visualization of the \dr results is used to draw a conclusion rather than generate a hypothesis.  The hypothesis is often that the high dimensional data has sufficient information to separate the variable of interest, often a binary or categorical variable.  This workflow is frequently used when there is a well-known phenomenon that is potentially expensive to measure.  An alternative method may be proposed to gather high-dimensional measurements of the object to be classified and then hypothesize that in this high-dimensional measurement, the data is separable.  This measurement may be a novel process (i.e. \phys~electric tongue~\cite{pauliuc2020raspberry}, Surface-enhanced Raman scattering sensors~\cite{yasukuni2019quantitative}), \chem (i.e. different types of spectrometry of liquids~\cite{zhou2020research, schievano2020nmr, suhandy2021use} or electric measurements of scent~\cite{wei2021novel, li2021accelerating}), \bio (i.e. microbiome data~\cite{xu2019synchronous}, MALDI fingerprinting~\cite{petukhova2018whole} or gene expression data~\cite{wei2021novel}), or \busi (i.e. novel qualitative analyses of businesses, countries, or groups of consumers~\cite{jonek2022assessing, lin2020exploring, malafronte2021integrated}).

\phys~Pauliuc et al.~\cite{pauliuc2020raspberry} provides an example where PCA is used to reduce data from a measurement of honey using a process called a \textit{Voltammetric Electronic Tongue} to determine which flower influenced its flavors.  In this common type of confirmatory data analysis workflow, the separability of clusters in the visualization (see Fig.~\ref{fig:workflows} (b)) is used as confirmation that the process can successfully recover the type of flower, or if the honey was not the product of a single type of flower but rather a combination of flowers, which is a less desirable type of honey.  The process being validated may also be a statistical process, as in \busi~\cite{llorente2023role}, where authors conduct a cluster analysis based on survey data analyzing internet habits of older populations in Spain.  A PCA view of the data is annotated with the convex hull of the discovered clusters, showing the separability of most of the clusters.

\begin{figure}[ht]
\centering
\includegraphics[width=\columnwidth]{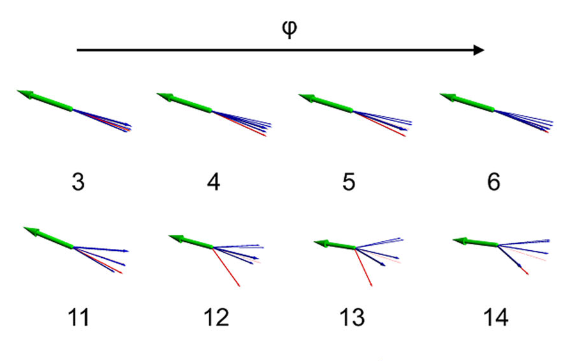}
    \caption{In Otten et al.~\cite{otten2021event}, the PCA scores are converted to polar coordinates, and points within that space are visualized as vectors, with the thickness of the green arrow and directions of the blue and red arrows encoding physically meaningful values in the original data.}
    \label{fig:phys_directions}
\end{figure}

A novel example of this workflow is found in \phys~Otten et al.~\cite{otten2021event}, where a deep generative model is used to generate events in a physical process.  The analytical goal is to interpret the data generated from the deep generative model, evaluating it for use in further analysis.  PCA is used to reduce the dimensionality of a hidden layer of the generative model to the first two principal components.  However, before visualizing the data, that two-dimensional space is converted to polar coordinates, and samples are taken on a grid in polar coordinates (Fig.~\ref{fig:phys_directions}).  Samples on this grid are given a visual encoding representing multiple dimensions of the underlying physical data.

Lastly, this type of workflow was used not only to confirm hypotheses but also to reject hypotheses.  \chem~Nurani et al. ~\cite{nurani2021metabolite} used H-NMR spectroscopy to extract metabolite data about turmeric plants in order to classify the particular species.  They use several \dr techniques to identify clusters and evaluate separability and conclude that PCA is only able to distinguish certain species and not others: "\textit{Chemometrics of PCA could not differentiate C. longa, C. xanthorrhiza, and C. manga clearly (data not shown). It might be caused by the large variations of the variables; therefore, the principal components (PC) were not able to represent the original variables.}"  Interestingly, they do not include a plot of the results they use to make this conclusion.  They go on to state that a different \drshort technique was able to distinguish between them: "\textit{Observation using supervised pattern recognition, namely PLS-DA using 7 PC, could classify C. longa, C. xanthorriza, and C. manga resulting in three different classifications.}"

It is likely that this type of workflow is broadly analogous to assessing the value of a projection by its perceptual cluster separability, which is a quantitative metric describing the distance and clarity of separation that has been identified as a quality metric for projection techniques~\cite{sedlmair2015data}.  Cluster separability on projections of data with known ground truth can be used in a quantitative evaluation or comparison of different projection techniques.


\noindent \textbf{Workflow 3: Confirmatory then Exploratory Data Analysis}  In the third workflow, a projection is generated to ultimately be used for exploratory data analysis to generate hypotheses for the correlation between data features.  However, in order to build a greater level of confidence in the projected view, it is first evaluated via confirmatory data analysis.  

This workflow is frequently seen in domains where there is complex input data being used to study a phenomenon that is not well understood.  First, the projection is inspected visually, with points colored according to some known quantity that should be separable within the data.  Visual separation confirms that there is some meaning in the layout of the points in the projected view.  Then, the projection is repeated but with additional data encoded on each point to develop new hypotheses and potentially enrich the understanding of the meaning of known separable groups.  The analysis then continues into additional exploration and confirmation of those hypotheses through other statistical tests.  

This type of workflow was typical in \bio~(eg. ~\cite{cui2020activation, potluri2020antibody, kosicki2022cas9, higa2022spatiotemporal, wang2022decoding}) where many physiological processes interact and are typically represented by a high dimensional dataset, and there is a large space of potential hypotheses that can be narrowed through exploration of high dimensional data.  As an example, in Cui et al. (see Fig.~\ref{fig:workflows}(c)), a tSNE projection of lung cells is first projected to confirm that abnormal tissue is separable from normal tissue, as well as cell type~\cite{cui2020activation}.  Then, that view is colored by the variables of interest, pJUN and pAKT, to understand if their concentration differs from abnormal tissue to normal tissue.  It is identified that fibrotic lung tissue sees high readings of both variables in a particular cell type, fibroblasts, which are then further analyzed.  This technique can sometimes be used to color many variables of interest, making use of small multiples (see \bio~\autoref{fig:plotstyle}b).

One notable example was found in \busi~Teng et al.~\cite{teng2019principal}.   A factory process is being optimized, and PCA is used to find a subset of processes that might be easily experimented on together without disrupting too much of the manufacturing process.  First, PCA scores for the first four components are shown against the high dimensional features and confirmed to find meaningful clusters, and then the clusters are used to drive the design of experiments and optimizations using domain knowledge of the factory itself.  This analysis is seen in \autoref{fig:plottypes}c.  

\begin{figure}[!tp]
\centering
\includegraphics[width=\columnwidth]{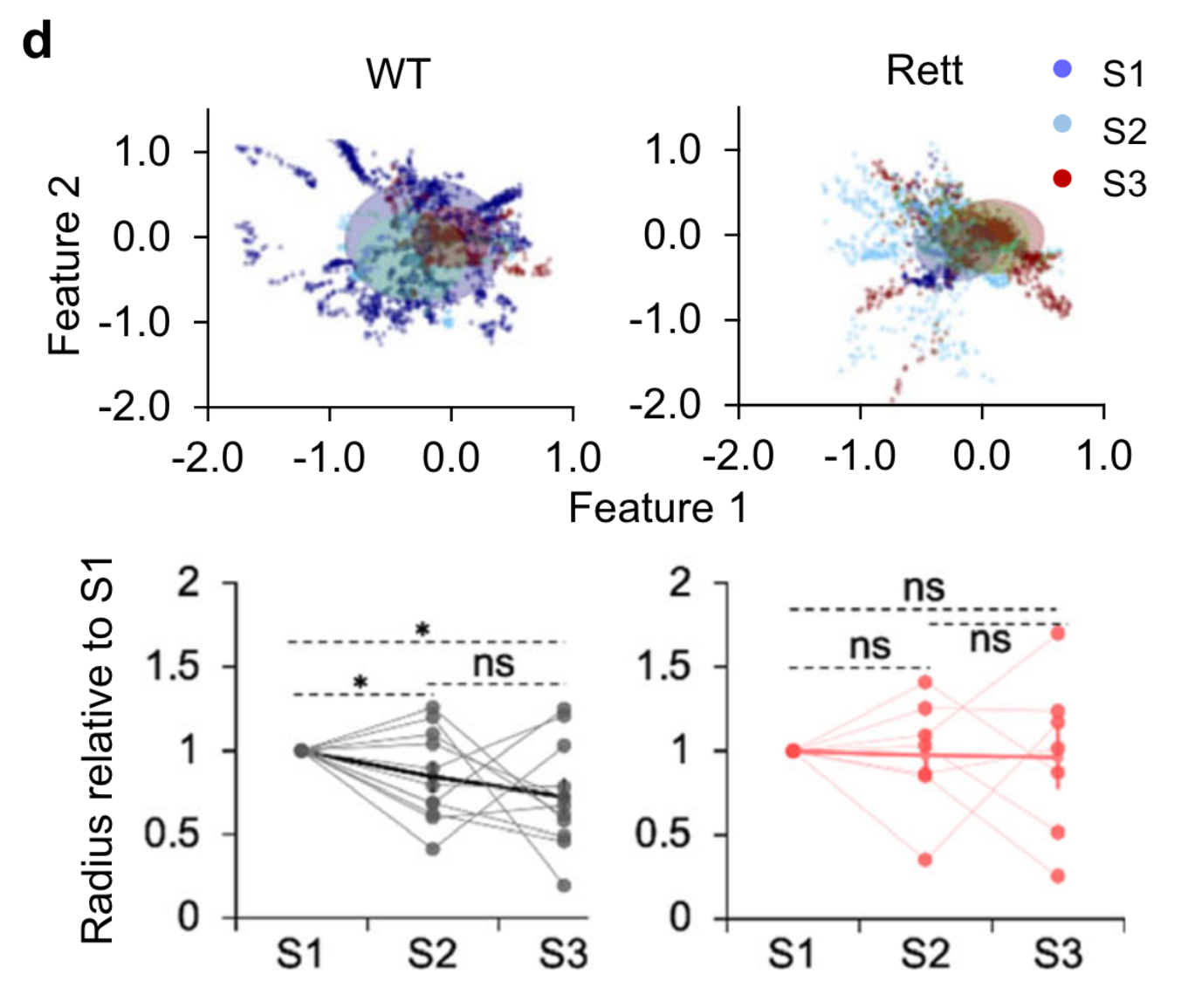}
    \caption{In Xu et al. \cite{xu2022pattern}, authors use a Variational Autoencoder (VAE) to reduce the dimensionality of in vivo image frames of mouse neurons. The authors recorded brain imaging data for two types of mice doing three types of tasks, S1, S2, and S3.  The linear distance from each point to the cluster center of points in task S1 is reported in the bottom plots. However, the VAE method can distort distances, making this possibly misleading information (more in \autoref{sec:incorrect_usages}).
    }
    \label{fig:xu2022pattern}
\end{figure}

\subsection{Gaps between Research and Data Analysis}
\label{sec:gaps}

Our literature review identified gaps between the practical use of \drshort and the relevant research conducted by the visualization community. 
The findings ignite open challenges for the visualization community to reduce the gap.

\subsubsection{Inappropriate Usages}

\label{sec:incorrect_usages}

\begin{figure*}[tbh!]
    \centering
    \includegraphics[width=\linewidth]{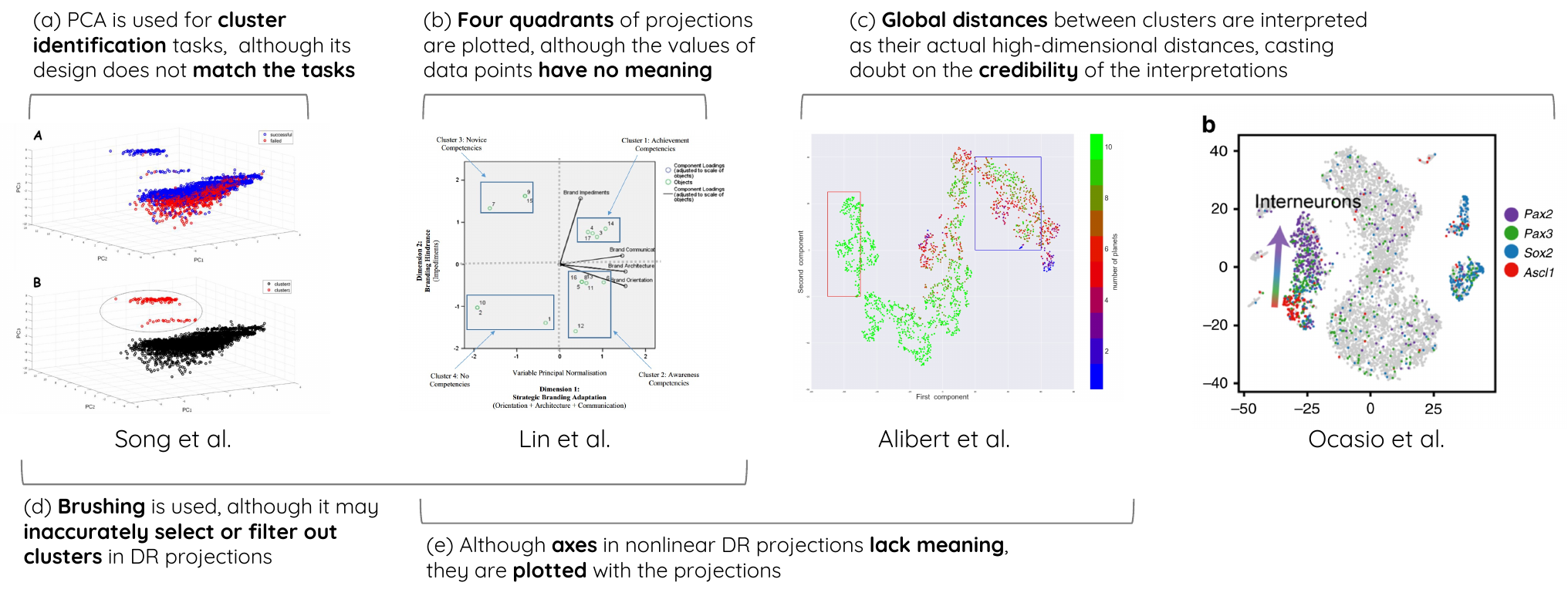}
    \caption{The examples of our findings on inappropriate usage of DR techniques (\autoref{sec:incorrect_usages}), found in \cite{song2019mining, lin2020exploring, alibert2019new, ocasio2019scrnaseq}. These improper usages lead data analysis to have limited reliability, casting doubt on the conclusions in which research works made. }
    \label{fig:incorrect_usages}
\end{figure*}

Our analysis reveals inappropriate usages of DR techniques that could lead to erroneous conclusions about the relying data, e.g., relying on assumptions that were not guaranteed by the methods being used. Please refer to \autoref{fig:incorrect_usages} for examples.

\noindent
\textbf{Inappropriate choice of DR techniques}
The visualization community has analyzed which \drshort techniques are best suited for various visual analytics tasks, such as cluster identification and neighborhood search. 
For example, Xia et al. \cite{xia22tvcg} reveal that UMAP and t-SNE are the most appropriate techniques for cluster identification tasks. 
They found that PCA is not suitable for the task, which means that PCA shows inaccurate cluster representations. 
This is largely achieved by conducting benchmark studies, where the appropriateness of DR techniques is evaluated using scores from DR quality metrics \cite{espadoto2021toward} or human task accuracy \cite{xia22tvcg, etemadpour15tvcg}. 

However, we identify that research works in four domains often use DR techniques that do not match with their task. We especially find that PCA is widely used for cluster identification tasks (\autoref{fig:incorrect_usages}a), although it is less suitable for the task \cite{xia22tvcg}. This inappropriate usage degrades the reliability of the findings made by the research work; for example, the clusters found by the practitioners may not stay as clusters in the high-dimensional space \cite{aupetit2007visualizing, aupetit14beliv}.

\noindent
\textbf{Inappropriate plotting of DR techniques}
The visualization community also informed practitioners how to plot \drshort projections properly. 
For example, Faust et al. \cite{faust19tvcg} emphasized that conventional $x$ and $y$ cannot be used to interpret nonlinear \drshort techniques, proposing a new nonlinear axes visualization technique. 
Jeon et al. \cite{jeon2022distortion} and Aupetit et al. \cite{aupetit14vast} claimed that conventional brushing that selects 2D regions should not be used for \drshort projections, contributing new brushing techniques that locally resolve distortions.

Still, we find cases where research works in four domains do not align with such guidelines.  
For axes plotting, we identify papers that plot axes with titles such as UMAP1 \& UMAP2 or TSNE1 \& TSNE2 or with grid lines (\autoref{fig:incorrect_usages}e). 
We also find the case in which four quadrants of the dimensionally reduced view are used to define four clusters (\autoref{fig:incorrect_usages}b), suggesting that the positive and negative directions of the axes hold semantic meaning for the authors.
Regardless, these annotations could mislead audiences, in particular, those unfamiliar with the properties of the underlying DR techniques used.


In terms of brushing, we find that several papers emphasize the clusters using fixed-shape brushes (e.g., rectangular, ellipse, or spheres), which violates the guidelines made by the visualization community (\autoref{fig:incorrect_usages}d). As with inappropriate technique selection, such a violation degrades the credibility of brushed clusters.

\noindent
\textbf{Inappropriate interpretation of data patterns}
The visualization and machine learning community provided guidelines to interpret DR projections based on the design of the DR techniques used. For example, Wattenberg et al. \cite{wattenberg2016use} guided practitioners not to interpret global distances between clusters in t-SNE plot as their distances in the original high-dimensional space. The claim has been further verified by many articles, not only for t-SNE but also for other nonlinear DR techniques like UMAP \cite{jeon24classes, fu2019atsne, jeon22uniform, moor2020topological, moon2019visualizing}.

However, our review reveals that research works in four domains often do not comply with such guidelines.
For example, it was common to interpret the global distances between clusters in DR projections that do not preserve the global structure (\autoref{fig:incorrect_usages}c), casting doubt on the credibility of the analysis results.

In another example, \phys~Xu et al. \cite{xu2022pattern} project frames of mouse brain imaging data into a 2D latent space using a Variational Autoencoder (VAE) (Fig. \ref{fig:xu2022pattern}).
The authors stratify the data points into groups based on experimental conditions.
For each group of points, they compute a distribution radius defined as ``\textit{the average distance of all frames to the center in the 2D
latent space.}''
The authors proceed to draw conclusions by comparing the distribution radii among experimental conditions.
Because VAE is a potentially non-linear method, it is possible for distances in the 2D space to be distorted, preventing linear comparison.

\noindent \textbf{Execution of statistical test using dimension-reduced data}
The biology and biostatistics community provided guidance not to conduct statistical tests using dimension-reduced data \cite{kriegeskorte2009circular,lahnemann2020eleven}. 
This is because the data is distorted during the reduction process and then fed to the statistical test, which means that it has been ``double-dipped.'' 

However, several research works in four domains apply statistical tests like the $t$-test to the PCA results, casting doubt on the validity of the test (e.g., \cite{song2019mining}). Even though the solutions for this problem have been widely proposed in the biology community \cite{gao2022selectiveinference,chen2022selectiveinference,neufeld2021treevalues,song2023scdesign3}, the incorrect execution of statistical tests is rampant in the field.

\subsubsection{Issues with Reproducibility}

\label{sec:reproducibility}

We find that in most cases, the authors specified the software package (e.g., \texttt{scikit-learn} \cite{pedregosa2011scikit}) that was used to perform \drshort and reported how they preprocessed data. 
However, papers failed to mention whether optional parameters (e.g., perplexity in the case of t-SNE) were used, degrading their
reproducibility. 
This is because most works have used the default hyperparameter settings of the library they used. Still, there are possibilities in which hyperparameters are cherry-picked to generate DR projections that best align with the papers' hypotheses. The absence of a hyperparameter report also raises concerns about whether the hyperparameter values have been properly optimized, negatively impacting not only reproducibility but also the credibility of the data analysis.

\section{Discussion}

\label{sec:discussion}

In this section, we summarize the takeaways of our survey for both target audiences: 1) visualization and machine learning researchers and 2) domain-centered data analysts and practitioners.
We note that due to our selection of four domains, there are some limitations to the generalizability of our findings, but we believe that the diversity of our four domains lends weight to our findings and, thus, our takeaways and open problems.


\subsection{Takeaways For Domain Researchers}

A theme of this survey is that there is a disconnection between DR-focused research in the visualization community and DR usage in the subject areas of Biology, Business, Chemistry, and Physics.
Despite a diversity in the types of findings that authors discuss when referencing DR plots, we find that most use 2D scatterplots for visualization.
Certain types of such findings may benefit from alternative visual encodings or may be apparent without the usage of dimensionality reduction.
We encourage researchers using DR methods to ask themselves what value the DR added to their analysis.
This way, the process of creating visualizations can be centered around the communication of those findings that are intrinsic to the usage of DR.

Domain scientists and visualization researchers should recognize the ways in which DR results can be interpreted and visualized correctly.
Solutions to problems discussed in Section \ref{sec:findings} have been proposed in certain subject areas, but researchers in others may not yet be aware.
For instance, den-SNE and densMAP published in Biology enable interpretation of point density that would be erroneous using the original t-SNE and UMAP algorithms \cite{narayan2020assessing}.
PHATE, also published in Biology, is a nonlinear DR method that preserves global structure and patterns, such as branch points and trajectories \cite{moon2019visualizing}.
Methods that account for double usage of data during statistical testing have been proposed in Biostatistics \cite{gao2022selectiveinference,chen2022selectiveinference,neufeld2021treevalues,song2023scdesign3}.


We hope this survey motivates further qualitative analyses of how dimensionality reduction techniques are used and visualized within and across fields.
Domain scientists may be able to provide DR usage guidance that is tailored to the data types and analysis workflows that are commonly encountered in the field.

\subsection{For Visualization Researchers}

We organize our takeaways for visualization researchers into two categories.  First, we provide takeaways into the usage of \drshort in visual analytics systems, and then focus on the need for developing new visual analytics techniques to address the needs of domain scientists.

\subsubsection{Mismatches in Usage Patterns}

There were many differences between the types of \dr used in our observed domain literature and that found in visualization research.  In a recent survey of the use of embeddings in visualization systems by Huang et al.~\cite{huang2023va+}, it was identified that most visual analytics systems use nonlinear \dr algorithms, such as t-SNE or UMAP.  Similarly,  according to a survey by Espadoto et al., 33 out of 44 (75\%) dimensionality reduction techniques used in visualization papers are nonlinear methods.  In our report, we find that PCA is overwhelmingly the most common technique.  Visual analytics designers could consider using linear techniques more frequently, because of their familiarity and ease of interpretation - especially if the phenomenon being studied is visible within the linear projected view or the number of data points is low.  Nonlinear techniques can be superior to linear techniques at cluster separability and anecdotally in identifying clusters in image databases.  However, 
Based on the workflows identified in section~\ref{sec:findings}, linear techniques may be more helpful in confirming or generating hypotheses.  



In addition, visual analytics designers should consider alternatives to scatterplots that better support their users' tasks.  If the goal is to understand differences between known clusters or categories in the data, it may be better to include other encodings of cluster distances or vector distances between data points.  Alternatively, visual analytics designers can consider interpretability as an objective in their choice of linear projection~\cite{gleicher2013explainers}.


Lastly, we note that out of all papers included in our report, none used the types of interactive analysis that are frequently featured in visual analytics systems for the type of cluster analysis frequently used on high-dimensional data~\cite{kwon2016axisketcher, kwon2017clustervision}.  Instead, domain scientists use standard techniques that are more easily reproducible.  
Creating easily usable open source software and publicizing methods for interpretability through tutorials or guidelines papers in meaningful venues for domains
could improve our ability to reach other communities.







\subsubsection{Research Opportunities for the Visualization Community}




\noindent
\textbf{Needs for a unified guideline that informs when to use which techniques}
The visualization field provided several guidelines for selecting a DR technique that matches analytic tasks \cite{nonato18multidimensional, xia22tvcg, sedlmair13tvcg}. Also, the community has provided several empirical studies that ground these guidelines \cite{xia22tvcg, espadoto2021toward, atzberger24tvcg}. However, as described in the recent survey by Espadoto et al.~\cite{espadoto2021toward}, these guidelines are fragmented, and the interaction techniques developed by the visual analytics community also broaden the ambiguity in selecting appropriate DR for a given context \dr~\cite{sacha2016visual}.
There is a need for simple guidelines on when to use what technique and how to incorporate interactive techniques without introducing bias.  This would mitigate the risk of early adopters within the domain sciences from using novel techniques.  An empirical guidelines work was published more than a decade ago by Bertini et al.~\cite{bertini2011quality}, but the landscape of \dr has changed enough that there is a demand for more guidelines.

\noindent
 \textbf{Tangible and detailed guidelines beyond the selection of techniques}
    This survey reveals that domain researchers often use DR techniques in ways that can lead to misinterpretations (\autoref{sec:incorrect_usages}), and their methodologies often lack reproducibility \autoref{sec:reproducibility}. These findings clearly indicate the need for more detailed guidelines that help domain researchers
    . As aforementioned, the visualization field provides several guidelines for using dimensionality reduction (DR), but these guidelines mostly map analytic tasks to techniques \cite{nonato18multidimensional, xia22tvcg}. They support domain researchers in selecting good techniques but do not offer insights on interpreting and communicating the projections. More tangible guidelines that fit domain researchers with lower visualization and machine learning literacy are thus needed. For example, a detailed protocol to comprehensively report DR execution and its results in their paper may substantially benefit in enhancing the reproducibility of DR-based visual analytics.  This challenge has been noted previously by \dr researchers~\cite{bibal2021biot, faust19tvcg, lambert2022globally}; based on the usage patterns found in our surveyed domain papers, we recommend visualization researchers build toolkits and target new venues for their design studies.

\noindent
\textbf{Packaging up our techniques into a toolkit} We generally found that domain scientists used DR techniques that were standard packages in common languages such as \textit{R} or \textit{Matlab}, or in some cases used languages developed for their particular type of data like viSNE~\cite{amir2013visne} or PHATE~\cite{moon2019visualizing}.
However, the visualization field currently lacks libraries that serve various techniques developed by a community. 
It is clear that we can improve the usability of these techniques by packaging them in clean interfaces available in package managers, which would also improve the likelihood that domain experts will use them \cite{jeon23zadu}.  
We also recommend that researchers make these packages more actionable. As Draco \cite{moritz2018formalizing} did for general visualization design, we can help domain experts by building a framework that automatically recommends DR techniques and hyperparameter settings that align with the experts' task and data domain.

\noindent 
    \textbf{Greater impact for design studies through targeted publishing}
    Our study indicated that in the four fields we have studied, there has not been penetration of best practices from our community.  We believe there is an opportunity for greater impact if we encourage authors of design studies to publish within the venues of the application domain, in addition to visualization domains.  By presenting the value of our approaches on datasets that are meaningful to domain scientists within their venues, we can make it easier for domain scientists to understand the value and risk of appropriate interpretations.  
    
    We also recommend that surveys of design studies also target applied domains.  While most surveys count domain scientists among the intended audience of their work, it may be unlikely that surveys in the visualization research community are ever encountered by domain scientists.  To mitigate this gap, we recommend that visualization researchers consider writing an executive summary of survey findings and share that within applied domains, potentially as letters or notes within their professional publications.
\section{Conclusion}
\label{sec:conclusion}
In this paper, we describe the state of the art in the usage and interpretation of dimensionality reduction in domain-specific data analysis across four domains: biology, chemistry, physics, and business.  We conduct three iterations of analysis: 1) a bibliometric analysis of papers citing dimensionality reduction techniques, 2) a loose analysis of papers using dimensionality reduction techniques, and 3) a structured analysis of papers across four domains from the last 5 years.  We classify their usage and interpretation of DR techniques and then describe qualitative findings.  We believe this study provides valuable insights to both domain scientists and computer science researchers in understanding the usage of tools and the gaps that could drive further research within the visualization community.

\bibliographystyle{IEEEtran}
\bibliography{reference}

\newpage

\begin{IEEEbiography}[{\includegraphics[width=1in,height=1.25in,clip,keepaspectratio]{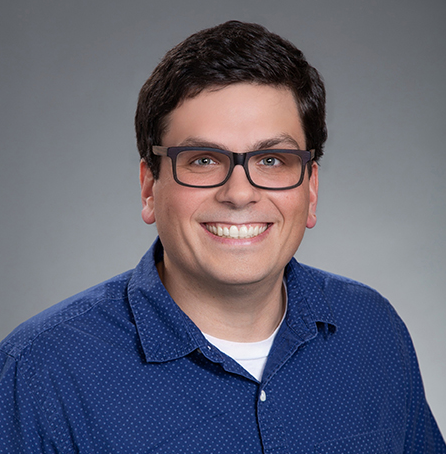}}]{Dylan Cashman}
is an assistant professor in the Michtom School of Computer Science at Brandeis University in Waltham, MA.  His research interests include the development and evaluation of visual affordances that improve usability of artificial intelligence models and data science processes. Dylan received a Ph.D in Computer Science from Tufts University. 
\end{IEEEbiography}

\begin{IEEEbiography}[{\includegraphics[width=1in,height=1.25in,clip,keepaspectratio]{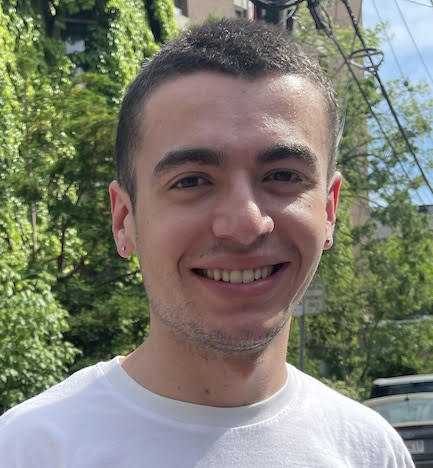}}]{Mark S. Keller}
is a student in the Bioinformatics and Integrative Genomics PhD Program at Harvard Medical School in Boston, MA. He holds a Bachelor of Science in Computer Science from University of Maryland, College Park. His research interests include the development of interactive visualization tools for high-dimensional single-cell omics data, including for transcriptomics and chromatin accessibility experiments.
\end{IEEEbiography}

\begin{IEEEbiography}[{\includegraphics[width=1in,height=1.25in,clip,keepaspectratio]{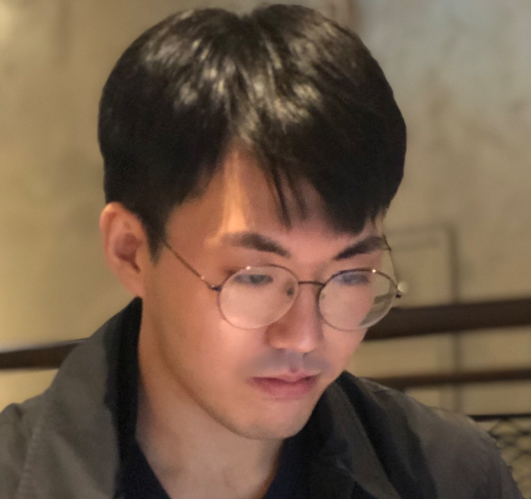}}]{Hyeon Jeon}
is a Ph.D student at Seoul National University, Seoul, Korea.  He is currently working on developing new visualizations and machine learning techniques that support reliable data analysis. Before starting his Ph.D. program, he received a B.S. degree in Computer Science and Engineering from the Pohang University of Science and Technology (POSTECH), Pohang, Korea. 
\end{IEEEbiography}

\begin{IEEEbiography}[{\includegraphics[width=1in,height=1.25in,clip,keepaspectratio]{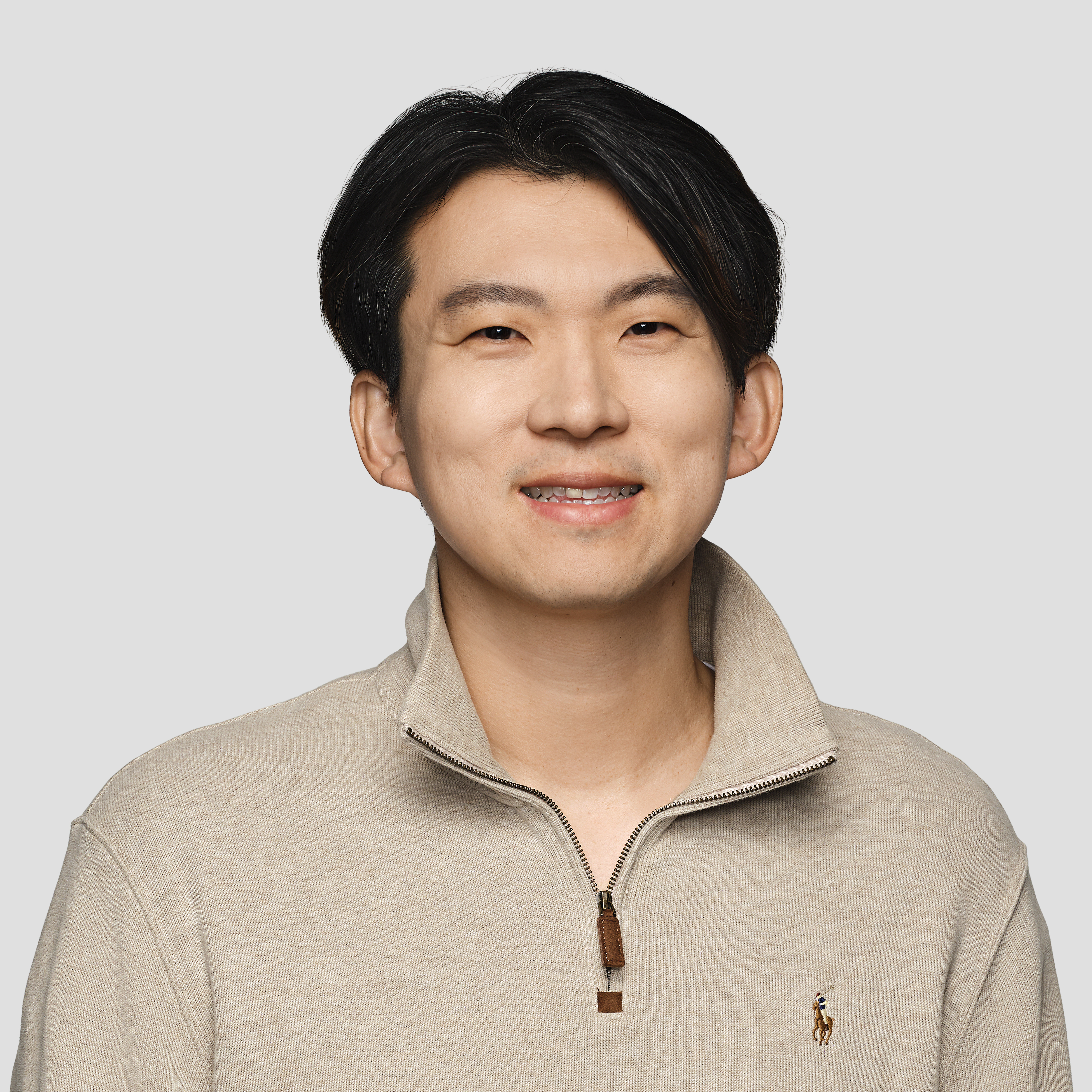}}]{Bum Chul Kwon}
is a researcher at IBM Research. His research area includes visual analytics, data visualization, human-computer interaction, healthcare, and machine learning. His primary research interest includes the development of interactive visualizations to enhance users' abilities to derive knowledge from biomedical data using interactive visualization systems.  He received his Ph.D in Data Visualization from Purdue University.
\end{IEEEbiography}

\begin{IEEEbiography}[{\includegraphics[width=1in,height=1.25in,clip,keepaspectratio]{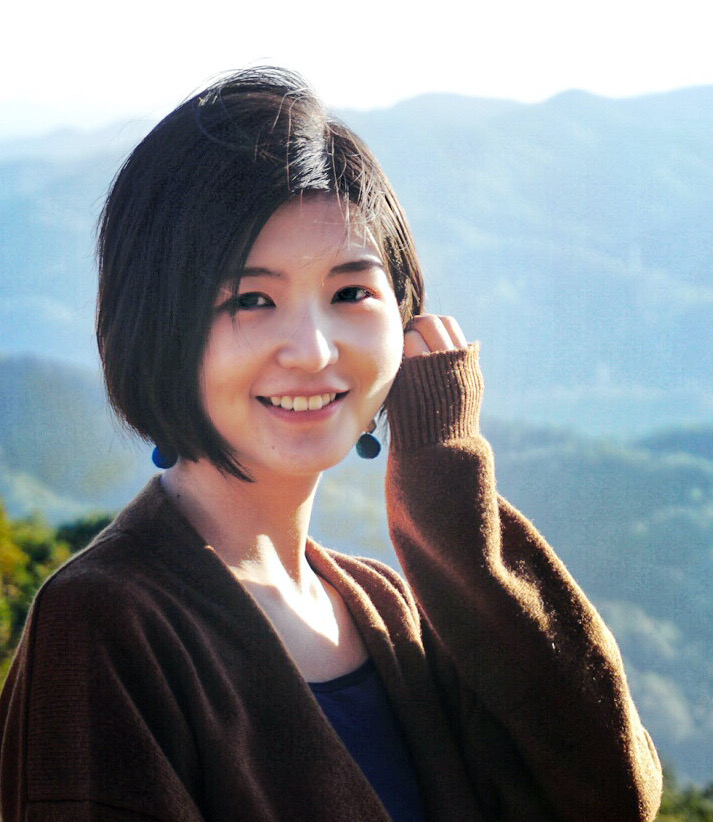}}]{Qianwen Wang}
is an Assistant Professor in the department of computer science and engineering at the University of Minnesota.  Her research aims to enhance communication and collaboration between domain users and AI through interactive visualizations, particularly focusing on their applications in addressing biomedical challenges. She received her Ph.D in Electronic and Computer Engineering from Hong Kong University of Science and Technology.
\end{IEEEbiography}








\newpage

\vfill

\end{document}